\documentclass{iopart}
\usepackage{iopams}

\newtheorem{Proposition}{Proposition}
\newtheorem{Lemma}{Lemma}
\newtheorem{Corollary}{Corollary}
\newtheorem{Definition}{Definition}

\newenvironment{proof}{\noindent{\bf proof.}}{\hspace{\stretch{1}}\opensquare}

\begin{document}

\title{Subspace local quantum channels}
\author{Johan \AA berg}
\address{Department of Quantum Chemistry, 
Uppsala University, Box 518, SE-751 20 Uppsala, Sweden}
\ead{johan.aaberg@kvac.uu.se}

\begin{abstract}
A special class of quantum channels, named \emph{subspace local} (SL),
are defined and investigated. The proposed definition of subspace
locality of quantum channels is an attempt to answer the question of
what kind of restriction should be put on a channel, if it is to act
`locally' with respect to two `locations', when these naturally
correspond to a separation of the total Hilbert space in an orthogonal
sum of subspaces $\mathcal{H} = \mathcal{H}_{1}\oplus\mathcal{H}_{2}$,
rather than a tensor product decomposition
$\mathcal{H}=\mathcal{H}_{1}\otimes\mathcal{H}_{2}$. It is shown that
the set of SL channels decomposes into four disjoint families of
channels. Explicit expressions to generate all channels in each family
are presented. It is shown that one of these four families, the
\emph{local subspace preserving} (LSP) channels, is precisely the
intersection between the set of \emph{subspace preserving} channels
and the SL channels. For a subclass of the LSP channels, a special
type of unitary representation using ancilla systems is presented.
\end{abstract}


\section{Introduction}
This investigation appears in a series that focuses on properties of
quantum channels with respect to orthogonal sum decompositions of
Hilbert spaces of quantum systems. The present study concentrates on
subspace locality of quantum channels. Subspace preserving CPMs are
investigated in \cite{ref1}, and the concept of gluing of CPMs in
\cite{ref3}.

Questions tied to locality is a central theme in many investigations
of quantum mechanics, like correlation, entanglement, teleportation,
and questions about what kind of nonlocal resources are needed in
performing various operations \cite{NilChu}. Usually a system
consisting of two separate entities on separate locations are modeled
with a Hilbert space in form of a tensor product
$\mathcal{H}_{1}\otimes\mathcal{H}_{2}$, where $\mathcal{H}_{1}$
represents the pure states of the subsystem at location $1$ and
correspondingly for $\mathcal{H}_{2}$. General states of the system
are represented by density operators on the Hilbert space. Operations
on this bipartite system is represented by `channels', which are trace
preserving completely positive maps (CPMs). Channels map states to
states of the system, and give a quite general tool to describe
evolution that allows interaction with external quantum systems. A
channel $\Phi$ can be said to be local with respect to this bipartite
system, if it can be written as a product channel
$\Phi=\Phi_{1}\otimes\Phi_{2}$, where $\Phi_{1}$ and $\Phi_{2}$ act on
density operators on the spaces $\mathcal{H}_{1}$ and
$\mathcal{H}_{2}$, respectively. In other words the total operation is
decomposed into operations which act only locally on each
subsystem. The product channels are precisely those which can be
achieved with `local means' only.

The central assumption in this modeling is that the separation in two
locations is associated with a decomposition of the total Hilbert
space into a tensor product. However, there are situations where the
separation in locations is not directly associated with such a tensor
decomposition, but rather a decomposition into two orthogonal
subspaces. One example is the two arms of a two path single particle
interferometer.
Reasonably, the Hilbert space $\mathcal{H}$ of the particle in the
interferometer is decomposed into an orthogonal sum of a Hilbert space
$\mathcal{H}_{1}$ representing the pure states localized in path $1$,
and a Hilbert space $\mathcal{H}_{2}$ representing pure states
localized in arm $2$. Hence, $\mathcal{H} =
\mathcal{H}_{1}\oplus\mathcal{H}_{2}$. Clearly the separation into the
two locations corresponds to a decomposition of the total Hilbert
space into an orthogonal sum.

Suppose we let a channel $\Phi$ act on the state of the particle in
the interferometer. In other words the particle can be affected in a
very general way, possibly interacting with other quantum systems,
while passing the arms of the interferometer.  What kind of
restrictions should be put on $\Phi$ if it is supposed to act
`locally' with respect to the two paths? Hence, $\Phi$ should be
possible to realize using only `local resources and means' on each
location. In other words, what we look for is a reasonable definition
of `local channel', when the separation into locations corresponds to
a decomposition into an orthogonal sum of the total Hilbert space.

The \emph{subspace local channels} presented here is an attempt for
such a definition.  Whatever definition one assumes, it should be
consistent with the definition of locality with respect to subsystems
i.e. separation into tensor product. The strategy used here to define
subspace locality, is to embed the original state space in a larger
state space, in such a way that the orthogonal sum decomposition is
`transformed' into a tensor decomposition.  When having this
decomposition in tensor product, the standard definition of locality
in terms of product channels can be applied. The embedding is achieved
by changing the basic description of the particle state, from being a
description of the state of the particle, to a description of the
occupation states at the two locations. In other words, an occupation
number representation of a second quantization of the original state
space. The definition of subspace local channels in terms of these
larger spaces is thereafter translated back to the original state
space.

The main advantage of this definition is that it is based on the
standard definition of product channels, in a way that can be given a
physical interpretation, and that can be applied in very general
settings. It must however be emphasized that it is not certain that
this is the most appropriate definition in all contexts. In this study
the consequences of this definition is investigated. Future
investigations will have to settle whether or not this, or other
alternative definitions, are best suited to fit the intuitive notion
of ``subspace local'' channels.

The approach to use a second quantized description to describe the
state of a single particle has been used in discussions of
single-particle non-locality. For examples see \cite{Tan},
\cite{Har1}, \cite{Bjork}, and \cite{BjH}.

The investigations performed here are all made under the assumption
that the Hilbert spaces involved are finite-dimensional. This
restriction is primarily made of practical reasons, to get reasonably
rigorous derivations without getting too involved into mathematical
technicalities. The author believes that many of the propositions,
with suitable technical modifications, remains true for
infinite-dimensional separable Hilbert spaces. That question will,
however, not be treated here.

The structure of this article is as follows: in section \ref{defnot}
notation and conventions are presented. In section \ref{mai} we give
the basic definition of subspace locality in terms of second
quantization of the state spaces of the systems. It is shown that a
large part of these second quantized spaces are irrelevant for the
analysis and that the definition can be reformulated on smaller
subspaces. In section \ref{indep} the local subspace preserving
channels (LSP) are defined. In section \ref{seunitrep} a special type
of unitary representation of a subclass of the LSP channels is
presented. In section \ref{seSL} we turn to the set of SL channels and
show that this set is partitioned into four families of
channels. Explicit formulas to generate all the channels in each of
these four families, is deduced.  In section \ref{dis} we discuss some
conceptual aspects of the nature of the SL channels.  A summary is
given in \ref{sum}.
\section{Notation and conventions}
\label{defnot}
Complex Hilbert spaces are denoted by $\mathcal{H}$ with various
subscripts.  Given two Hilbert spaces $\mathcal{H}_{S}$ and
$\mathcal{H}_{T}$, a CPM $\phi$ is a linear map from the set of
trace-class operators $\tau(\mathcal{H}_{S})$ to the set of
trace-class operators $\tau(\mathcal{H}_{T})$. We say that
$\mathcal{H}_{S}$ is the \emph{source space} of $\phi$ (or just
\emph{source}), and $\mathcal{H}_{T}$ is the \emph{target space} of
$\phi$ (or just \emph{target}). On finite-dimensional Hilbert spaces,
the set of trace class operators coincide with the set of linear
operators. We let $\mathcal{L}(\mathcal{H})$ denote the set of all
linear operators from the Hilbert space $\mathcal{H}$ to itself and
let $\mathcal{L}(\mathcal{H}_{S},\mathcal{H}_{T})$ denote the set of
linear operators from $\mathcal{H}_{S}$ to $\mathcal{H}_{T}$. Trace
preserving CPMs (channels) are denoted by capital Greek letters, to
distinguish them from the general CPMs which are denoted by small
Greek letters.

Orthogonal decompositions in pairs of subspaces of both the source and
target space, i.e.  $\mathcal{H}_{S} =
\mathcal{H}_{s1}\oplus\mathcal{H}_{s2}$ and $\mathcal{H}_{T} =
\mathcal{H}_{t1}\oplus\mathcal{H}_{t2}$, play a central role. These
subspaces are assumed to be at least one-dimensional, hence
$\mathcal{H}_{S}$ and $\mathcal{H}_{T}$ are both at least
two-dimensional. To each of these spaces correspond orthogonal
projectors and orthonormal bases. Each row of the following table
consists of a space, the corresponding projection operator, and the
notation for an arbitrary orthonormal basis of the subspace in
question:
\begin{displaymath} 
\begin{array}{ccc}
\mathcal{H}_{S} & \mathcal{H}_{s1}  & \mathcal{H}_{s2} \\
\hat{1}_{S} & P_{s1} & P_{s2}\\
\{|S,n\rangle\}_{n} & \{|s1,k\rangle\}_{k} & \{|s2,l\rangle\}_{l},
\end{array}
\end{displaymath}
where the index span the appropriate number of elements in each case.
Similar notation is used for the spaces $\mathcal{H}_{T}$,
$\mathcal{H}_{t1}$, and $\mathcal{H}_{t2}$. The spaces
$\mathcal{H}_{T}$, $\mathcal{H}_{S}$, and their subspaces are referred
to as the `first quantized' spaces.

Unfortunately the list of spaces does not end here but continues with
second quantized versions of these spaces. $F^{x}(\mathcal{H})$
denotes the occupation number representation of a second quantization
of the Hilbert space $\mathcal{H}$. The index $x$ denotes the type of
second quantization (fermionic or bosonic). This choice determines how
many particles can occupy a state in the first quantized
space. $F_{1}(\mathcal{H})$ represents the pure single-particle
states.  $F_{0}(\mathcal{H})$ is the space spanned by the vacuum
state, and $F_{01}(\mathcal{H}) = F_{0}(\mathcal{H})\oplus
F_{1}(\mathcal{H})$. These spaces do not carry any index to denote the
type of quantization, since these subspaces are independent of that
choice. $F^{x}_{2}(\mathcal{H})$ represents all pure states with at
least two particles. This subspace is affected by the choice of
statistics.  One may note that $F_{1}(\mathcal{H}) \simeq \mathcal{H}$
and that $F_{0}(\mathcal{H})\simeq \mathbb{C}$.  If $\mathcal{H}$ is
one-dimensional and if we have chosen fermionic second quantization,
then $F^{x}_{2}(\mathcal{H})$ is zero-dimensional. The following table
shows the notation for the relevant subspaces of the second
quantization of the source space.
\begin{displaymath} 
\begin{array}{cccc}
F_{0}(\mathcal{H}_{S}) & F_{1}(\mathcal{H}_{S}) &
F_{01}(\mathcal{H}_{S}) & F^{x}_{2}(\mathcal{H}_{S})\\
P_{\widetilde{S}:0} & P_{\widetilde{S}:1} & P_{\widetilde{S}:01} &
P_{\widetilde{S}:2}\\
\{|\widetilde{0}_{S}\rangle\} & \{|\widetilde{S}:1,n\rangle\}_{n} & 
\{|\widetilde{0}_{S}\rangle\}\cup\{|\widetilde{S}:1,n\rangle\}_{n} & 
\{|\widetilde{S}:2,m\rangle\}_{m}
\end{array}
\end{displaymath}
We use an analogous notation for the second quantizations of the
target space. Second quantizations of the subspaces
$\mathcal{H}_{s1}$, $\mathcal{H}_{s2}$, $\mathcal{H}_{t1}$, and
$\mathcal{H}_{t2}$ are also used, with the following notation:
\begin{displaymath} 
\begin{array}{cccc}
F_{0}(\mathcal{H}_{s1}) & F_{1}(\mathcal{H}_{s1}) &
F_{01}(\mathcal{H}_{s1}) & F^{x}_{2}(\mathcal{H}_{s1}) \\
P_{\widetilde{s1}:0} & P_{\widetilde{s1}:1} & P_{\widetilde{s1}:01} &
P_{\widetilde{s1}:2} \\
\{|\widetilde{0}_{s1}\rangle\} & \{|\widetilde{s1},k\rangle\}_{k}  & 
\{|\widetilde{0}_{s1}\rangle\}\cup\{|\widetilde{s1},k\rangle\}_{k}  & 
\{|\widetilde{s1}:2,m\rangle\}_{m} 
\end{array}
\end{displaymath}
Similarly for $\mathcal{H}_{s2}$, $\mathcal{H}_{t1}$, and
$\mathcal{H}_{t2}$.  To make more clear when an object (vector,
operator, CPM) `belongs' to a second quantized space it carries a
tilde, while objects belonging to a first quantized space do not.

\emph{Restriction in source space} and \emph{restriction in target space} 
of CPMs have been defined in \cite{ref3}. Given a subspace
$\mathcal{H}_{s1}$ in the source space of a CPM, the \emph{restriction
in source space} of $\phi$, to $\mathcal{H}_{s1}$ (or just
\emph{restriction in source}), is defined as the restriction of $\phi$
to the subspace $\mathcal{L}(\mathcal{H}_{s1})$. Given a subspace
$\mathcal{H}_{t1}$ of the target space of a CPM $\phi$ the
\emph{restriction in target space} is defined as $\chi(Q) =
P_{t1}\phi(Q)P_{t1}$ for all $Q\in\mathcal{L}(\mathcal{H}_{S})$. It is
to be noted that there is a slight abuse of notation in the last
expression. The mapping $\chi$ is intended to have $\mathcal{H}_{t1}$ as its
target space, not $\mathcal{H}_{T}$.  The facts to be used here is
that the restriction in source or target of a CPM, is a CPM. Moreover,
the restriction in source space of a trace preserving CPM is trace
preserving \cite{ref3}.

Given a CPM $\widetilde{\phi}$ with source $F^{x}(\mathcal{H}_{S})$
and target $F^{x'}(\mathcal{H}_{T})$, define the \emph{1-restriction}
of $\widetilde{\phi}$ to be the CPM which is the restriction in source
to $F_{1}(\mathcal{H}_{S})$ and restriction in target to
$F_{1}(\mathcal{H}_{T})$, of $\widetilde{\phi}$. To handle
1-restrictions it will prove convenient to use the following
operators:
\begin{eqnarray*}
M_{S} & =
\sum_{k}|\widetilde{s1},k\rangle|\widetilde{0}_{s2}\rangle\langle s1,k|
+\sum_{l}|\widetilde{0}_{s1}\rangle|\widetilde{s2},l\rangle\langle s2,l|,
\\  M_{T} & 
=\sum_{n}|\widetilde{t1},n\rangle|\widetilde{0}_{t2}\rangle\langle t1,n| +
\sum_{m}|\widetilde{0}_{t1}\rangle|\widetilde{t2},m\rangle\langle t2,m|,
\end{eqnarray*}
\begin{eqnarray*}
M_{s1}  = \sum_{k}|\widetilde{s1},k\rangle\langle s1,k|, & 
\quad M_{s2}  = \sum_{l}|\widetilde{s2},l\rangle\langle s2,l|, \\
M_{t1}  = \sum_{n}|\widetilde{t1},n\rangle\langle t1,n|, & 
\quad M_{t2}  = \sum_{m}|\widetilde{t2},m\rangle\langle t2,m|. 
\end{eqnarray*}
In terms of these operators the 1-restriction of a CPM
$\widetilde{\phi}$ can be expressed as
\begin{equation}
\label{reduc}
\phi(Q) = M_{T}^{\dagger}\widetilde{\phi}(M_{S}QM_{S}^{\dagger})M_{T}.
\end{equation}
 
An operator $V:\mathcal{H}_{s1}\rightarrow \mathcal{H}_{t1}$ will in
some expressions be treated as a mapping from $\mathcal{H}_{S}$ to
$\mathcal{H}_{T}$. In such cases it is implicitly assumed that $V$
acts as the zero operator on $\mathcal{H}_{s2}$, and that $V$ is
linearly extended to $\mathcal{H}_{S}$. The range of the original $V$
is some subspace of $\mathcal{H}_{t1}$, but we instead regard it as a
subspace of $\mathcal{H}_{T}$. This convention is for example used in
equation (\ref{pr}) in proposition \ref{islekv}.  Another notational
simplification, in the same spirit as the previous one, concerns
CPMs. Given a CPM $\phi$ with source $\mathcal{H}_{s1}$ and target
$\mathcal{H}_{t1}$, it will in some cases be treated as if it had
source space $\mathcal{H}_{S}$ and target $\mathcal{H}_{T}$. Given a
Kraus representation $\{V_{k}\}_{k}$ of $\phi$ we re-interpret $V_{k}$
as being mappings from $\mathcal{H}_{S}$ to $\mathcal{H}_{T}$, as
mentioned above. This new set of operators forms a Kraus
representation of a CPM with source space $\mathcal{H}_{S}$ and target
space $\mathcal{H}_{T}$.  A typical example of this convention is 
$\phi = \phi_{1}+\phi_{2}$, where $\phi_{1}$ has source
$\mathcal{H}_{s1}$ and target $\mathcal{H}_{t1}$, and $\phi_{2}$ has
source $\mathcal{H}_{s2}$ and target $\mathcal{H}_{t2}$.

A final remark concerns linear maps with $\mathcal{L}(\mathcal{H})$ as
its domain of definition. If a linear map $\Lambda$ fulfills
$\Lambda(|\psi\rangle\langle\psi|)=0$ for all
$|\psi\rangle\in\mathcal{H}_{s1}$, then this statement is equivalent
to: $\Lambda(\rho)=0$ for all density operators $\rho$ on
$\mathcal{H}_{s1}$, and is moreover equivalent to: $\Lambda(Q)=0$,
$\forall Q\in\mathcal{L}(\mathcal{H}_{s1})$. This is the case, since
every density operator can be written as a convex combination of outer
products of elements in $\mathcal{H}$, and since every
$Q\in\mathcal{L}(\mathcal{H}_{s1})$ can be written as a (complex)
linear combination of four density operators. We will in the following
pass between these equivalent formulations without any comment.
\section{Definition of subspace locality}
\label{mai}
We begin with a brief description of some of the main ideas appearing
in this section.  The key observation is the following:
$F^{x}(\mathcal{H}_{1}\oplus\mathcal{H}_{2})=
F^{x}(\mathcal{H}_{1})\otimes F^{x}(\mathcal{H}_{2})$. Hence, when
passing to the second quantization in the occupation number
representation, the original orthogonal sum gives rise to a tensor
product. With respect to this tensor product the subspace local
channels can be defined as those which correspond to product channels
on these product spaces.

To use channels on the second quantized spaces to represent channels
on the first quantized spaces, introduce some problems.  When passing
from the first quantized description to a second quantized
description, we allow more operations, as one may change the total
particle number in the system. If an operation on the second quantized
space is to act as a channel on the first quantized space, we must
make sure that initial single-particle states are mapped to
single-particle states. The channels which \emph{respects 1-states}
are those for which there corresponds a channel on the first quantized
spaces.  The second problem is that to a specific channel on first
quantized spaces there correspond several possible channels on the second
quantized spaces. The \emph{1-restriction} singles out those channels
which correspond to a specific channel on the first quantized spaces.
A third problem is that the type of second quantization is not
unique. Although one may argue, from an intuitive point of view, that
the type of second quantization should not matter, since we are
dealing with single particles, this has to be proved. Proposition
\ref{reduprod} shows that the type of second quantization indeed does
not matter. It also shows that, for the questions dealt with here,
only parts of the second quantized spaces are relevant and that the
definition of subspace locality can be reformulated in terms of these
subspaces.
\begin{Definition}\rm   
Let the CPM $\widetilde{\phi}$ have source $F^{x}(\mathcal{H}_{S})$
and target $F^{x'}(\mathcal{H}_{T})$. We say that $\widetilde{\phi}$
\emph{respects 1-states} if
$\Tr(P_{\widetilde{T}:1}^{\perp}\widetilde{\phi}
(|\widetilde{\psi}\rangle\langle\widetilde{\psi}|)) = 0$, $\forall
|\widetilde{\psi}\rangle\in F_{1}(\mathcal{H}_{S})$, where
$P_{\widetilde{T}:1}^{\perp}$ denotes the projector onto the
orthogonal complement of $F_{1}(\mathcal{H}_{T})$.  Likewise
$\widetilde{\phi}$ \emph{respects 0-states} if
$\Tr(P_{\widetilde{T}:0}^{\perp}\widetilde{\phi}
(|\widetilde{0}_{S}\rangle\langle\widetilde{0}_{S}|))
= 0$, where $P_{\widetilde{T}:0}^{\perp}$ is the projector onto the
orthogonal complement of $F_{0}(\mathcal{H}_{T})$.  If
$\widetilde{\phi}$ both respects 1-states and 0-states we say that
$\widetilde{\phi}$ \emph{respects 1,0-states}.
\end{Definition}
In words, a CPM on a second quantized space respects 1-states if no
single-particle state is mapped outside the set of strict
single-particle states. (We make no restriction on how
$\widetilde{\phi}$ may act on states outside the strict
single-particle states.)  These definitions will be used also for CPMs
with source $F_{01}(\mathcal{H}_{S})$ and target
$F_{01}(\mathcal{H}_{T})$.
\begin{Lemma}
\label{lere}
Let $\widetilde{\phi}$ be a CPM with source $F^{x}(\mathcal{H}_{S})$
and target $F^{x'}(\mathcal{H}_{T})$, or alternatively with source
$F_{01}(\mathcal{H}_{S})$ and target $F_{01}(\mathcal{H}_{T})$. If
$\widetilde{\phi}$ respects n-states ($n= 0$ or $1$), then
$P_{\widetilde{T}:n}\widetilde{\phi}(\widetilde{Q})P_{\widetilde{T}:n}
= \widetilde{\phi}(\widetilde{Q})$ for all $\widetilde{Q}\in
\mathcal{L}(F_{n}(\mathcal{H}_{S}))$.
\end{Lemma}
\begin{proof}
Let $\{\widetilde{V}_{k}\}_{k}$ be some Kraus representation of
$\widetilde{\phi}$. If $\widetilde{\phi}$ respects n-states it follows
by definition that
$\sum_{k}\Tr(P_{\widetilde{T}:n}^{\perp}
\widetilde{V}_{k}|\widetilde{\psi}\rangle
\langle\widetilde{\psi}|\widetilde{V}_{k}^{\dagger})
= 0$, $\forall |\widetilde{\psi}\rangle\in F_{n}(\mathcal{H}_{S})$. 
It follows that
$\langle\widetilde{\chi}|\widetilde{V}_{k}|\widetilde{\psi}\rangle =
0$ for all $k$, all $|\widetilde{\chi}\rangle\in
F_{n}(\mathcal{H}_{T})^{\perp}$, and all $|\widetilde{\psi}\rangle\in
F_{n}(\mathcal{H}_{S})$. Hence,
$P_{\widetilde{T}:n}^{\perp}\widetilde{V}_{k}P_{\widetilde{S}:n}=0$,
from which the lemma follows.
\end{proof}

To single out those channels on the second quantized systems that
correspond to channels on the first quantized system, the following
lemma is useful, since it gives us the choice to either show trace
preservation of the 1-restriction, or to show that the channel on the
second quantized system respects 1-states.
\begin{Lemma}
\label{retr}
Let $\widetilde{\Phi}$ be a trace preserving CPM with source
$F^{x}(\mathcal{H}_{S})$ and target $F^{x'}(\mathcal{H}_{T})$, or
alternatively with source $F_{01}(\mathcal{H}_{S})$ and target
$F_{01}(\mathcal{H}_{T})$. Then $\widetilde{\Phi}$ respects 1-states
if and only if the 1-restriction is trace preserving.
\end{Lemma}
\begin{proof}
By assuming that the 1-restriction is trace preserving, it follows
that
$\Tr(M_{T}^{\dagger}\widetilde{\Phi}(M_{S}|\psi\rangle
\langle\psi|M_{S}^{\dagger})M_{T})
=1$, for all normalized $|\psi\rangle\in\mathcal{H}_{S}$. By combining
$M_{T}M_{T}^{\dagger} = P_{\widetilde{T}:1}$,
$P_{\widetilde{T}:1}+P_{\widetilde{T}:1}^{\perp} =\hat{1}$ and the
assumption that $\widetilde{\Phi}$ is trace preserving, follows 
$\Tr(P_{\widetilde{T}:1}^{\perp}\widetilde{\Phi}(M_{S}
|\psi\rangle\langle\psi|M_{S}^{\dagger}))
=0$. Since $M_{S}$ is a bijection between $\mathcal{H}_{S}$ and
$F_{1}(\mathcal{H}_{S})$, it follows from the last expression that
$\widetilde{\Phi}$ respects 1-states. Reversing the argument above
gives that if $\widetilde{\Phi}$ respects 1-states, then the
1-restriction is trace preserving.
\end{proof}

We are now in position to state the definition of subspace locality of
channels.
\begin{Definition}\rm
Let $\Phi$ be a trace preserving CPM with source
$\mathcal{H}_{s1}\oplus\mathcal{H}_{s2}$ and target
$\mathcal{H}_{t1}\oplus\mathcal{H}_{t2}$. If there exist trace
preserving CPMs $\widetilde{\Phi}_{1}$ and $\widetilde{\Phi}_{2}$,
where $\widetilde{\Phi}_{1}$ has source $F^{x}(\mathcal{H}_{s1})$ and
target $F^{x'}(\mathcal{H}_{t1})$, and where $\widetilde{\Phi}_{2}$
has source $F^{x}(\mathcal{H}_{s2})$ and target
$F^{x'}(\mathcal{H}_{t2})$, for some choice of $x,x'$, such that
$\widetilde{\Phi}_{1}\otimes\widetilde{\Phi}_{2}$ has $\Phi$ as
1-restriction, then we say that $\Phi$ is \emph{subspace local} from
$(\mathcal{H}_{s1},\mathcal{H}_{s2})$ to
$(\mathcal{H}_{t1},\mathcal{H}_{t2})$.
\end{Definition} 
In words this definition says that a CPM $\Phi$ is subspace local from
$(\mathcal{H}_{s1},\mathcal{H}_{s2})$ to
$(\mathcal{H}_{t1},\mathcal{H}_{t2})$ if there exist some second
quantizations of the source space and target space, and some product
channel $\widetilde{\Phi}_{1}\otimes\widetilde{\Phi}_{2}$, such that
$\widetilde{\Phi}_{1}\otimes\widetilde{\Phi}_{2}$ `acts' like $\Phi$
on the single-particle states. Hence, we use the familiar definition
of `local' channel, to define what should be meant by subspace local
channel.
   
It will now be shown that a large part of the second quantized space
is irrelevant to the analysis of SL channels. A consequence is that
the set of subspace local channels is independent of the choice of
statistics in the second quantization.
 
Let $|\widetilde{\psi}_{1}\rangle$ be an arbitrary normalized state in
$F_{01}(\mathcal{H}_{t1})$.  Consider the CPM
$\widetilde{\Theta}_{1}$, with source and target
$F^{x'}(\mathcal{H}_{t1})$, defined by
\begin{eqnarray}
\label{etadef}
\fl\widetilde{\Theta}_{1}(\widetilde{Q}_{1}) =  
P_{\widetilde{t1}:01}\widetilde{Q}_{1} P_{\widetilde{t1}:01} + 
\sum_{l}|\widetilde{\psi}_{1}\rangle
\langle\widetilde{t1}:2, l| \widetilde{Q}_{1}|\widetilde{t1}:2, l\rangle
 \langle \widetilde{\psi}_{1}|\nonumber\\
\lo= P_{\widetilde{t1}:01}\widetilde{Q}_{1}P_{\widetilde{t1}:01} +
 |\widetilde{\psi}_{1}\rangle
\langle \widetilde{\psi}_{1}|\Tr(P_{\widetilde{t1}:2} \widetilde{Q}_{1}),
\quad\forall \widetilde{Q}_{1}\in\mathcal{L}(F^{x'}(\mathcal{H}_{t1})). 
\end{eqnarray}
In case $F^{x'}_{2}(\mathcal{H}_{t1})$ is zero-dimensional, discard
the second term in (\ref{etadef}).  By the form of
$\widetilde{\Theta}_{1}$ it follows that it is a CPM and moreover a
trace preserving CPM.  Construct analogously a trace preserving CPM
$\widetilde{\Theta}_{2}$, with source and target space
$F^{x'}(\mathcal{H}_{t2})$, using a normalized state
$|\widetilde{\psi}_{2}\rangle\in F_{01}(\mathcal{H}_{t2})$.
\begin{Lemma}
\label{lehjalp}
Let $\widetilde{\phi} =
\widetilde{\phi}_{1}\otimes\widetilde{\phi}_{2}$, where the CPM
$\widetilde{\phi}_{1}$ has source $F^{x}(\mathcal{H}_{s1})$ and target
$F^{x'}(\mathcal{H}_{t1})$, and where the CPM $\widetilde{\phi}_{2}$
has source $F^{x}(\mathcal{H}_{s2})$ and target
$F^{x'}(\mathcal{H}_{t2})$. Let $\widetilde{\phi}'
=\widetilde{\phi}_{1}'\otimes\widetilde{\phi}_{2}' $ be defined by
$\widetilde{\phi}_{1}' =
\widetilde{\Theta}_{1}\circ\widetilde{\phi}_{1}$,
$\widetilde{\phi}_{2}' =
\widetilde{\Theta}_{2}\circ\widetilde{\phi}_{2}$.
\begin{itemize}
\item  If  $\widetilde{\phi}$ respects 1-states then so does 
$\widetilde{\phi}'$.
\item  If  $\widetilde{\phi}$ respects 1-states, then 
$\widetilde{\phi}$ and $\widetilde{\phi}'$ have the same 1-restriction.
\item  If $\widetilde{\phi}$ is trace preserving, 
then so is $\widetilde{\phi}'$.
\end{itemize}
\end{Lemma}
\begin{proof}
We note the following:
$(\langle\widetilde{t1}:2,l|\langle\widetilde{t2}:2,l'|)P_{\widetilde{T}:1}
= 0$ for every $l,l'$,
$(P_{\widetilde{t1}:01}\otimes|\widetilde{\psi}_{2}\rangle
\langle\widetilde{t2}:2,
l'|)P_{\widetilde{T}:1} = 0$ for every $l'$,
$(|\widetilde{\psi}_{1}\rangle\langle\widetilde{t1}:2, l|\otimes
P_{\widetilde{t2}:01})P_{\widetilde{T}:1} = 0$ for every $l$, while
$(P_{\widetilde{t1}:01}\otimes
P_{\widetilde{t2}:01})P_{\widetilde{T}:1} = P_{\widetilde{T}:1}$. It
follows that
$[\widetilde{\Theta}_{1}\otimes\widetilde{\Theta}_{2}]
(P_{\widetilde{T}:1}\widetilde{Q}P_{\widetilde{T}:1})
= P_{\widetilde{T}:1}\widetilde{Q}P_{\widetilde{T}:1}$.

By combining the last expression with lemma \ref{lere} and the
definition of respect of 1-states, the first part of the lemma
follows.

For the second part one can use that $M_{S}|\psi\rangle\in
F_{1}(\mathcal{H}_{S})$, for any $|\psi\rangle\in\mathcal{H}_{S}$. 
By lemma \ref{lere} follows
$M_{T}^{\dagger}\widetilde{\phi}'
(M_{S}|\psi\rangle\langle\psi|M_{S}^{\dagger})M_{T}
= M_{T}^{\dagger}
[\widetilde{\Theta}_{1}\otimes\widetilde{\Theta}_{2}]
\left(P_{\widetilde{T}:1}\phi(M_{S}|\psi\rangle\langle\psi|M_{S}^{\dagger})
P_{\widetilde{T}:1}\right) =
M_{T}^{\dagger}\widetilde{\phi}(M_{S}|\psi\rangle\langle\psi|M_{S}^{\dagger})$,
which proves the second part of the lemma.

For the third part one can use that both $\widetilde{\Theta}_{1}$ and
$\widetilde{\Theta}_{2}$ are trace preserving. Hence,
$\widetilde{\Theta}_{1}\otimes\widetilde{\Theta}_{2}$ is trace
preserving. From this follows that $\widetilde{\phi}'$ is trace
preserving if $\widetilde{\phi}$ is.
\end{proof}
\begin{Proposition}
\label{reduprod}
Let $\Phi$ be a trace preserving CPM with source
$\mathcal{H}_{s1}\oplus\mathcal{H}_{s2}$ and target
$\mathcal{H}_{t1}\oplus\mathcal{H}_{t2}$. $\Phi$ is SL from
$(\mathcal{H}_{s1},\mathcal{H}_{s2})$ to
$(\mathcal{H}_{t1},\mathcal{H}_{t2})$, if and only if there exists a
trace preserving CPM $\widetilde{\Psi} = \widetilde{\Psi}_{1}\otimes
\widetilde{\Psi}_{2}$, where $\widetilde{\Psi}_{1}$ has source
$F_{01}(\mathcal{H}_{s1})$ and target $F_{01}(\mathcal{H}_{t1})$, and
where $\widetilde{\Psi}_{2}$ has source $F_{01}(\mathcal{H}_{s2})$ and
target $F_{01}(\mathcal{H}_{t2})$, such that $\widetilde{\Psi}$ has
$\Phi$ as 1-restriction.
\end{Proposition}
\begin{proof}
We begin with the ``only if'' part of the proposition. Suppose $\Phi$
is subspace local from $(\mathcal{H}_{s1},\mathcal{H}_{s2})$ to
$(\mathcal{H}_{t1},\mathcal{H}_{t2})$. By definition there exists a
trace preserving product CPM
$\widetilde{\Phi}=\widetilde{\Phi}_{1}\otimes\widetilde{\Phi}_{2}$,
which has $\Phi$ as its 1-restriction. If the construction in lemma
\ref{lehjalp} is applied, a product CPM $\widetilde{\Phi}'
=\widetilde{\Phi}'_{1} \otimes\widetilde{\Phi}'_{2}$ is
obtained. Lemma \ref{lehjalp} gives that $\widetilde{\Phi}'$ is trace
preserving and has $\Phi$ as its 1-restriction. Let
$\widetilde{\Psi}_{1}$ be the restriction in source to
$F_{01}(\mathcal{H}_{s1})$, and in target to
$F_{01}(\mathcal{H}_{t1})$, of $\widetilde{\Phi}'_{1}$. Let
$\widetilde{\Psi}_{2}$ be the restriction in source to
$F_{01}(\mathcal{H}_{s2})$, and in target to
$F_{01}(\mathcal{H}_{t2})$, of $\widetilde{\Phi}'_{2}$. Since
$F_{1}(\mathcal{H}_{S})\subset F_{01}(\mathcal{H}_{s1})\otimes
F_{01}(\mathcal{H}_{s2})$ and $F_{1}(\mathcal{H}_{T})\subset
F_{01}(\mathcal{H}_{t1})\otimes F_{01}(\mathcal{H}_{t2})$, it follows
that the 1-restriction of
$\widetilde{\Psi}=\widetilde{\Psi}_{1}\otimes\widetilde{\Psi}_{2}$ is
the same as the 1-restriction of $\widetilde{\Phi}'$.  Finally it
follows from the construction of $\widetilde{\Phi}'_{1}$ and
$\widetilde{\Phi}'_{2}$, that $\widetilde{\Psi}_{1}$ and
$\widetilde{\Psi}_{2}$ are trace preserving.

We turn to the ``if-part''. $\widetilde{\Psi}_{1}$ and $
\widetilde{\Psi}_{2}$ are trace preserving CPMs such that
$\widetilde{\Psi} =\widetilde{\Psi}_{1}\otimes \widetilde{\Psi}_{2}$
has $\Phi$ as 1-restriction. Assume $F^{x}_{2}(\mathcal{H}_{s1})$ is
not zero-dimensional and let the CPM $\widetilde{\eta}_{1}$ be defined
by $\widetilde{\eta}_{1}(\widetilde{Q}_{1}) =
\sum_{l}|\widetilde{0}_{t1}\rangle\langle
\widetilde{s1},2:l|\widetilde{Q}_{1}|\widetilde{s1},2:l\rangle\langle
\widetilde{0}_{t1}|$, $\forall
\widetilde{Q}_{1}\in\mathcal{L}(F^{x}(\mathcal{H}_{s1}))$.  Define
$\widetilde{\Phi}_{1} = \widetilde{\Psi}_{1} + \widetilde{\eta}_{1}$.
By construction it follows that $\widetilde{\Phi}_{1}$ is a trace
preserving CPM. If $F^{x}_{2}(\mathcal{H}_{s1})$ is zero-dimensional,
the term $\widetilde{\eta}_{1}$ drops out. In analogy with
$\widetilde{\eta}_{1}$, a CPM $\widetilde{\eta}_{2}$ can be
constructed, from which we define $\widetilde{\Phi}_{2}
=\widetilde{\Psi}_{2} + \widetilde{\eta}_{2}$.  If
$\widetilde{\Psi}_{1}\otimes\widetilde{\Psi}_{2}$ has $\Phi$ as
1-restriction, so does
$\widetilde{\Phi}_{1}\otimes\widetilde{\Phi}_{2}$. Hence, $\Phi$ is
subspace local.
\end{proof}  
\section{Local subspace preserving channels}
\label{indep}
Here a special class of SL channels is defined, the \emph{local
subspace preserving} (LSP) channels. Intuitively these are the
channels which are subspace local and which do not transport any
probability weight from one location to the other. It is shown in
section \ref{seSL} that the set of LSP channels is the intersection
between the set of subspace preserving channels (SP) \cite{ref1} and
the set of SL channels.
\begin{Definition}\rm
Let $\Phi$ be a trace preserving CPM with source
$\mathcal{H}_{s1}\oplus\mathcal{H}_{s2}$ and target
$\mathcal{H}_{t1}\oplus\mathcal{H}_{t2}$. We say that $\Phi$ is
\emph{local subspace preserving} (LSP) from
$(\mathcal{H}_{s1},\mathcal{H}_{s2})$ to
$(\mathcal{H}_{t1},\mathcal{H}_{t2})$ if there exists a trace
preserving CPM $\widetilde{\Phi}_{1}$ with source
$F_{01}(\mathcal{H}_{s1})$ and target $F_{01}(\mathcal{H}_{t1})$, and
a trace preserving CPM $\widetilde{\Phi}_{2}$ with source
$F_{01}(\mathcal{H}_{s2})$ and target $F_{01}(\mathcal{H}_{t2})$, such
that $\widetilde{\Phi}_{1}\otimes\widetilde{\Phi}_{2}$ has $\Phi$ as
1-restriction and such that both $\widetilde{\Phi}_{1}$ and
$\widetilde{\Phi}_{2}$ respects 1,0-states.
\end{Definition}
The following proposition gives an explicit description of the set of
LSP channels. It does so without referring to any second quantized
space. In other words, the definition of LSP channels as formulated in
the second quantized spaces, is `brought back' to the first quantized
spaces.
\begin{Proposition}
\label{islekv}
A trace preserving CPM $\Phi$ is LSP from
$(\mathcal{H}_{s1},\mathcal{H}_{s2})$ to
$(\mathcal{H}_{t1},\mathcal{H}_{t2})$ if and only if there exists a
linearly independent Kraus representation $\{V_{n}\}_{n=1}^{N}\subset
\mathcal{L}(\mathcal{H}_{t1},\mathcal{H}_{s1})$ of some trace
preserving CPM with source $\mathcal{H}_{s1}$ and target
$\mathcal{H}_{t1}$, and a linearly independent Kraus representation
$\{W_{m}\}_{m=1}^{M}\subset
\mathcal{L}(\mathcal{H}_{t2},\mathcal{H}_{s2})$ of some trace
preserving CPM with source $\mathcal{H}_{s2}$ and target
$\mathcal{H}_{t2}$, such that
\begin{equation}
\label{pr}
\fl\Phi(Q)= \sum_{n=1}^{N}V_{n}QV_{n}^{\dagger} + 
\sum_{m=1}^{M}W_{m}QW_{m}^{\dagger}+ VQW^{\dagger} + WQV^{\dagger},
\quad\forall Q\in\mathcal{L}(\mathcal{H}_{S}),
\end{equation} 
\begin{equation}
\label{VW}
V = \sum_{n=1}^{N} c_{1,n}V_{n},\quad W = \sum_{m =1}^{M}c_{2,m}W_{m},
\end{equation} 
where the vectors $c_{1} = [c_{1,n}]_{n=1}^{N}$ and $c_{2} =
[c_{1,m}]_{m=1}^{M}$ fulfill the conditions
\begin{equation}
\label{kondvar1}
||c_{1}||^{2} = \sum_{n=1}^{N}|c_{1,n}|^{2} \leq 1,\quad ||c_{2}||^{2}
= \sum_{m=1}^{M}|c_{2,m}|^{2} \leq 1.
\end{equation}
\end{Proposition}
This proposition is stated in terms of two linearly independent Kraus
representations. There exists another equivalent formulation of this
proposition in terms of arbitrary bases of
$\mathcal{L}(\mathcal{H}_{s1},\mathcal{H}_{t1})$ and
$\mathcal{L}(\mathcal{H}_{s2},\mathcal{H}_{t2})$. This is analogous to
the relation between proposition 10 in \cite{ref1} and proposition 3
in \cite{ref3}. The alternative formulation of proposition
\ref{islekv} can be obtained from proposition 10 in \cite{ref1} by
adding the condition that the matrix $C$ should be possible to write
as $C = c_{1}c_{2}^{\dagger}$, for some vector $c_{1}\in
\mathbb{C}^{K}$ and some $c_{2}\in\mathbb{C}^{L}$. Moreover, a
condition for trace preservation must be added, which takes the form
$\sum_{kk'}A_{k,k'}V_{k'}^{\dagger}V_{k}=P_{s1}$ and
$\sum_{ll'}B_{l,l'}W_{l'}^{\dagger}W_{l}=P_{s2}$. (Here the notation
in proposition 10 in \cite{ref1} has been used.)

\begin{proof}
By definition, $\Phi$ is an LSP channel if and only if there exist
trace preserving CPMs $\widetilde{\Phi}_{1}$ and
$\widetilde{\Phi}_{2}$, such that
$\widetilde{\Phi}_{1}\otimes\widetilde{\Phi}_{2}$ has $\Phi$ as
1-restriction, and such that both $\widetilde{\Phi}_{1}$ and
$\widetilde{\Phi}_{2}$ respects 1,0-states. Since
$\widetilde{\Phi}_{1}$ has source $F_{01}(\mathcal{H}_{s1})$ and
target $F_{01}(\mathcal{H}_{t1})$, the condition that
$\widetilde{\Phi}_{1}$ respects 1,0-states is equivalent to
$\widetilde{\Phi}_{1}$ being SP from
$(F_{0}(\mathcal{H}_{s1}),F_{1}(\mathcal{H}_{s1}))$ to
$(F_{0}(\mathcal{H}_{s2}),F_{1}(\mathcal{H}_{s2}))$ (compare with
definition of SP in \cite{ref1}). According to proposition 1 in
\cite{ref3}, $\widetilde{\Phi}_{1}$ is SP if and only if it is a
trace preserving gluing of two trace preserving CPMs. One of these
having source $F_{0}(\mathcal{H}_{s1})$ and target
$F_{0}(\mathcal{H}_{t1})$, and the other having source
$F_{1}(\mathcal{H}_{s1})$ and target $F_{1}(\mathcal{H}_{t1})$. The
spaces $F_{0}(\mathcal{H}_{s1})$ and $F_{0}(\mathcal{H}_{t1})$ are
one-dimensional and there exists only one trace preserving CPM with
these as source and target spaces, namely the CPM with linearly
independent Kraus representation
$\{|\widetilde{0}_{t1}\rangle\langle\widetilde{0}_{s1}|\}$.  The other
trace preserving CPM, with source $F_{1}(\mathcal{H}_{s1})$ and target
$F_{1}(\mathcal{H}_{t1})$, has some linearly independent Kraus
representation $\{\widetilde{V}_{n}\}_{n=1}^{N}$. According to
proposition 3 in \cite{ref3}, $\widetilde{\Phi}_{1}$ is an SP
channel, if and only if it can be written as
\begin{eqnarray}
\label{phi1}
\widetilde{\Phi}_{1}(\widetilde{Q}_{1}) = 
&\sum_{n}\widetilde{V}_{n}\widetilde{Q}_{1}\widetilde{V}_{n}^{\dagger} + 
|\widetilde{0}_{t1}\rangle
\langle \widetilde{0}_{s1}|\widetilde{Q}_{1}|\widetilde{0}_{s1}\rangle
\langle \widetilde{0}_{t1}|\nonumber \\
 &+\sum_{n}c_{1,n}\widetilde{V}_{n}\widetilde{Q}_{1}|\widetilde{0}_{s1}\rangle
\langle\widetilde{0}_{t1}| +
 \sum_{n}c_{1,n}^{*}|\widetilde{0}_{t1}\rangle\langle
 \widetilde{0}_{s1}|\widetilde{Q}_{1}\widetilde{V}_{n}^{\dagger},
\end{eqnarray}
for all $\widetilde{Q}_{1}\in\mathcal{L}(F_{01}(\mathcal{H}_{s1}))$,
where $c_{1} = [c_{1,n}]_{n=1}^{N}$ fulfills the condition
$||c_{1}||^{2} = \sum_{n=1}^{N}|c_{1,n}|^{2} \leq 1$.  By an analogous
reasoning, $\widetilde{\Phi}_{2}$ is an SP channel if and only if
\begin{eqnarray}
\label{phi2}
\widetilde{\Phi}_{2}(\widetilde{Q}_{2}) = 
&\sum_{m}\widetilde{W}_{m}\widetilde{Q}_{2}\widetilde{W}_{m}^{\dagger} + 
|\widetilde{0}_{t2}\rangle
\langle \widetilde{0}_{s2}|\widetilde{Q}_{2}|\widetilde{0}_{s2}\rangle
\langle \widetilde{0}_{t2}|\nonumber \\ 
&+\sum_{m}c_{2,m}\widetilde{W}_{m}\widetilde{Q}_{2}|\widetilde{0}_{s2}\rangle
\langle\widetilde{0}_{t2}| +
\sum_{m}c_{2,m}^{*}|\widetilde{0}_{t2}\rangle\langle
\widetilde{0}_{s2}|\widetilde{Q}_{2}\widetilde{W}_{m}^{\dagger},
\end{eqnarray}
for all $\widetilde{Q}_{2}\in\mathcal{L}(F_{01}(\mathcal{H}_{s2}))$,
where $c_{2} = [c_{2,m}]_{m=1}^{M}$ fulfills the condition
$||c_{2}||^{2} = \sum_{m=1}^{M}|c_{2,m}|^{2} \leq 1$, and where
$\{\widetilde{W}_{m}\}_{m=1}^{M}$ is a linearly independent Kraus
representation of some channel with source $F_{1}(\mathcal{H}_{s2})$
and target $F_{1}(\mathcal{H}_{t2})$.

The ``only if'' part of the proposition is now possible to prove by
taking the 1-restriction of
$\widetilde{\Phi}_{1}\otimes\widetilde{\Phi}_{2}$ using (\ref{reduc}).
Doing so one obtains expression (\ref{pr}), with $V_{n} =
M_{t1}^{\dagger}\widetilde{V}_{n}M_{s1}$ and $W_{m} =
M_{t2}^{\dagger}\widetilde{W}_{m}M_{s2}$. It is possible to check that
$\{V_{n}\}_{n=1}^{N}$ is a linearly independent Kraus representation
\cite{ref1} of a trace preserving CPM with source $\mathcal{H}_{s1}$
and target $\mathcal{H}_{t1}$. (One can make use of
$M_{s1}M_{s1}^{\dagger} = P_{\widetilde{s1}:1}$,
$M_{t1}M_{t1}^{\dagger} = P_{\widetilde{t1}:1}$, and
$P_{\widetilde{t1}:1}\widetilde{V}_{n} P_{\widetilde{s1}:1}
=\widetilde{V}_{n}$.) Analogously one can show that
$\{W_{m}\}_{m=1}^{M}$ is a linearly independent Kraus representation
of a channel with source $\mathcal{H}_{s2}$ and target
$\mathcal{H}_{t2}$.

For the ``if'' part, define $\widetilde{\Phi}_{1}$ and
$\widetilde{\Phi}_{2}$ via (\ref{phi1}) and (\ref{phi2}), with
$\widetilde{V}_{n} = M_{t1}V_{n}M_{s1}^{\dagger}$ and
$\widetilde{W}_{m} = M_{t2}W_{m}M_{s2}^{\dagger}$. Both
$\widetilde{\Phi}_{1}$ and $\widetilde{\Phi}_{2}$ respect 1,0-states
and $\widetilde{\Phi}_{1}\otimes\widetilde{\Phi}_{2}$ has $\Phi$ as
1-restriction.
\end{proof}
\subsection{Unitary representation of a subclass of the LSP channels}
\label{seunitrep}
A special class of LSP channels are those which have identical source
and target spaces and moreover where the separation in subspaces in
the source and target are
identical. $\mathcal{H}_{T}=\mathcal{H}_{S}$,
$\mathcal{H}_{t1}=\mathcal{H}_{s1}$ and
$\mathcal{H}_{t2}=\mathcal{H}_{s2}$. An example is a particle in a
two-path interferometer. To ease terminology a bit we say that a CPM
which is SP, LSP, or SL from $(\mathcal{H}_{s1},\mathcal{H}_{s2})$ to
$(\mathcal{H}_{s1},\mathcal{H}_{s2})$, is SP, LSP, or SL \emph{on}
$(\mathcal{H}_{s1},\mathcal{H}_{s2})$.

Here the channels in this subclass are constructed as a joint unitary
evolution of the system and an ancilla system. It is well known
\cite{Kraus} that any channel with identical source and target spaces
can be constructed via a joint unitary evolution. Here we prove a
special form of such a unitary representation.
\begin{Proposition}
\label{ISLunit}
A mapping $\Phi$ is LSP on $(\mathcal{H}_{s1},\mathcal{H}_{s2})$ if
and only if there exist finite-dimensional Hilbert spaces
$\mathcal{H}_{a1}$ and $\mathcal{H}_{a2}$ and normalized vectors
$|a1\rangle\in\mathcal{H}_{a1}$, $|a2\rangle\in\mathcal{H}_{a2}$, an
operator $V_{1}$ on $\mathcal{H}_{S}\otimes\mathcal{H}_{a1}$, and
$V_{2}$ on $\mathcal{H}_{S}\otimes\mathcal{H}_{a2}$, such that
\begin{equation}
\label{kondv}
V_{1}V_{1}^{\dagger} = V_{1}^{\dagger}V_{1} =
P_{s1}\otimes\hat{1}_{a1},\quad V_{2}V_{2}^{\dagger} =
V_{2}^{\dagger}V_{2} = P_{s2}\otimes\hat{1}_{a2},
\end{equation}  
and such that $\Phi$ can be written
\begin{equation}
\label{reprU1}
\Phi(Q) = \Tr_{a1,a2}\left(UQ\otimes|a1\rangle\langle a1|
\otimes|a2\rangle\langle a2|U^{\dagger}\right),\quad
\forall Q\in\mathcal{L}(\mathcal{H}_{S}),
\end{equation}
where $U = V_{1}\otimes\hat{1}_{a2} + V_{2}\otimes\hat{1}_{a1}$ is a
unitary operator.
\end{Proposition}
To have some intuition about this representation, note that the
unitary operator $U$ is decomposed into two parts. One part which acts
non-trivially only on the subspace $\mathcal{H}_{s1}$ and ancilla
$a1$, and another part which only acts on the subspace
$\mathcal{H}_{s2}$ and the second ancilla $a2$. Initially the two
ancilla systems are in a product state. Hence, we have a `local
resource' in form of two uncorrelated ancilla systems, one at each
location. In some sense, this evolution is such that it only acts
`locally', since if the state of the system is in subspace
$\mathcal{H}_{s1}$, it only interacts with ancilla system $a1$, and if
in subspace $\mathcal{H}_{s2}$, it only interacts with ancilla
$a2$. This seems to give some intuitive support to the here suggested
definition of subspace locality as being a reasonable definition.  A
more clearcut example is given in \cite{refinterf}, where the
expression (\ref{reprU1}) is applied to the special case of a two-path
single particle interferometer of particles with internal degrees of
freedom. In this special case the above proposition takes a more
transparent form, with unitary operators associated with the two paths
of the interferometer and acting on local ancillary systems. This
situation applies for example to investigations like \cite{sjomar} and
\cite{Oi}.

We may compare proposition \ref{ISLunit} with proposition 11 in
\cite{ref1}. There it is shown that, under the same restrictions on
the source and target spaces, there is a unitary representation for SP
channels. The expression derived there is similar to the expression in
proposition \ref{ISLunit}. The difference is that for the SP channels
there is only \emph{one} ancilla system and not two as in the case of
LSP channels. In the case of SP channels, both the `locations'
interact with the same ancilla system, which perhaps gives some
support to the idea that a general SP channel should not be regarded
as being `subspace local'.

\begin{proof}
Suppose $\Phi$ is LSP, then there exists a product channel of trace
preserving CPMs $\widetilde{\Phi}_{1}\otimes\widetilde{\Phi}_{2}$,
with $\Phi$ as 1-restriction. Consider $\widetilde{\Phi}_{1}$, which
has source $F_{01}(\mathcal{H}_{s1})$ and target
$F_{01}(\mathcal{H}_{s1})$. Since $\Phi$ is LSP it follows that
$\widetilde{\Phi}_{1}$ is trace preserving and SP on
$(F_{0}(\mathcal{H}_{s1}),F_{1}(\mathcal{H}_{s1}))$. By proposition 11
in \cite{ref1}, there exists a finite-dimensional Hilbert space
$\mathcal{H}_{a1}$, a normalized $|a1'\rangle\in\mathcal{H}_{a1}$, and
operators $\widetilde{W}_{1,0}$ and $\widetilde{W}_{1,1}$ on
$F_{01}(\mathcal{H}_{s1})\otimes\mathcal{H}_{a1}$, such that
\begin{equation}
\label{partis1}
\widetilde{W}_{1,0}\widetilde{W}_{1,0}^{\dagger}  = 
 \widetilde{W}_{1,0}^{\dagger}\widetilde{W}_{1,0} =
 |\widetilde{0}_{s1}\rangle\langle \widetilde{0}_{s1}|\otimes\hat{1}_{a1},
\end{equation}
\begin{equation}
\label{partis2}
\widetilde{W}_{1,1}\widetilde{W}_{1,1}^{\dagger}  = 
 \widetilde{W}_{1,1}^{\dagger}\widetilde{W}_{1,1} = 
P_{\widetilde{s1}:1}\otimes\hat{1}_{a1},
\end{equation}
\begin{equation}
\label{unitrep}
\widetilde{\Phi}_{1}(\widetilde{Q}_{1}) = 
\Tr_{a1}(\widetilde{U}'_{1} \widetilde{Q}_{1}\otimes|a1'\rangle\langle a1'|
 \widetilde{U}_{1}^{'\dagger}),\quad
 \forall \widetilde{Q}_{1}\in \mathcal{L}(F_{01}(\mathcal{H}_{s1})),
\end{equation}
where $\widetilde{U}'_{1} = \widetilde{W}_{1,0}+\widetilde{W}_{1,1}$
is a unitary operator.  Since $F_{0}(\mathcal{H}_{s1})$ is
one-dimensional, equation (\ref{partis1}) implies $\widetilde{W}_{1,0}
= |\widetilde{0}_{s1}\rangle\langle \widetilde{0}_{s1}|\otimes
U_{a1}$, where $U_{a1}$ is a unitary operator on
$\mathcal{H}_{a1}$. Define $\widetilde{V}_{1,0}$ by
\begin{displaymath}
\widetilde{V}_{1,0}= \widetilde{W}_{1,0}(\hat{1}\otimes U_{a1}^{\dagger})= 
|\widetilde{0}_{s1}\rangle\langle \widetilde{0}_{s1}|\otimes\hat{1}_{a1}.
\end{displaymath}
Define $\widetilde{V}_{1,1} = \widetilde{W}_{1,1}(\hat{1}\otimes
U_{a1}^{\dagger})$, $|a1\rangle = U_{a1}|a1'\rangle$, and
$\widetilde{U}_{1} =
\widetilde{V}_{1,0}+\widetilde{V}_{1,1}$. Equation (\ref{unitrep})
still holds with $\widetilde{U}'_{1}$ changed into $\widetilde{U}_{1}$
and $|a1'\rangle$ changed into $|a1\rangle$. Moreover, (\ref{partis1})
and (\ref{partis2}) imply
\begin{equation}
\label{WtV}
\fl \widetilde{V}_{1,1}\widetilde{V}_{1,1}^{\dagger}= 
\widetilde{V}_{1,1}^{\dagger}\widetilde{V}_{1,1} = 
P_{\widetilde{s1}:1}\otimes\hat{1}_{a1},\quad 
\widetilde{V}_{2,1}\widetilde{V}_{2,1}^{\dagger}= 
\widetilde{V}_{2,1}^{\dagger}\widetilde{V}_{2,1} =
 P_{\widetilde{s2},1}\otimes\hat{1}_{a2}.
\end{equation}
The operator $\widetilde{U}_{1}$ is unitary since $\widetilde{U}_{1}
=\widetilde{U'}_{1}(\hat{1}\otimes U_{a1})$.

An analogous argument can be applied to $\widetilde{\Phi}_{2}$, which
results in a representation of
$\widetilde{\Phi}_{1}\otimes\widetilde{\Phi}_{2}$ as
\begin{equation}
\label{eqPfU}
\fl [\widetilde{\Phi}_{1}\otimes\widetilde{\Phi}_{2}](\widetilde{Q}) =
 \Tr_{a1,a2}((\widetilde{U}_{1}\otimes \widetilde{U}_{2}) \widetilde{Q}
\otimes|a1\rangle\langle a1|\otimes
|a2\rangle\langle a2| 
(\widetilde{U}_{1}^{\dagger}\otimes \widetilde{U}_{1}^{\dagger})),
\end{equation}
where
\begin{equation}
\label{Udef}
\widetilde{U}_{1} = \widetilde{V}_{1,1} + 
|\widetilde{0}_{s1}\rangle\langle \widetilde{0}_{s1}|\otimes
\hat{1}_{a1},\quad \widetilde{U}_{2} = \widetilde{V}_{2,1} + 
|\widetilde{0}_{s2}\rangle\langle \widetilde{0}_{s2}|\otimes\hat{1}_{a2}.
\end{equation}
The next step is to calculate the 1-restriction of the channel
(\ref{eqPfU}), using (\ref{reduc}) (with $M_{T}=M_{S}$). One can use
that conditions (\ref{WtV}) imply
$(P_{\widetilde{s1}:1}\otimes\hat{1}_{a1})\widetilde{V}_{1,1}
(P_{\widetilde{s1}:1}\otimes\hat{1}_{a1})
= \widetilde{V}_{1,1}$. This can be proved using a singular value
decomposition \cite{LanTis} of $V_{1,1}$. (For more details see the
proof of proposition 11 in \cite{ref1}.)  The result of the
1-restriction is equation (\ref{reprU1}), with
\begin{equation}
\fl V_{1} = (M_{s1}^{\dagger}\otimes\hat{1}_{a1}) 
\widetilde{V}_{1,1}(M_{s1}\otimes\hat{1}_{a1}),\quad V_{2} = 
(M_{s2}^{\dagger}\otimes\hat{1}_{a2})\widetilde{V}_{2,1}
(M_{s2}\otimes\hat{1}_{a2}).
\end{equation}
One can check that the conditions (\ref{WtV}) imply (\ref{kondv}), and
from that showing that $U$ is unitary.

To prove the ``if part'' of the proposition, define
\begin{equation}
\fl\widetilde{V}_{1,1} = (M_{s1}\otimes\hat{1}_{a1})V_{1}
(M_{s1}^{\dagger}\otimes\hat{1}_{a1}),
\quad \widetilde{V}_{2,1} = 
(M_{s2}\otimes\hat{1}_{a2})V_{1}(M_{s2}^{\dagger}\otimes\hat{1}_{a2}).
\end{equation}
Define $\widetilde{U}_{1}$ and $\widetilde{U}_{2}$ via (\ref{Udef})
and define the CPM $\widetilde{\Phi}_{1}\otimes\widetilde{\Phi}_{2}$
via (\ref{eqPfU}). (This is a product CPM by construction). One can
check that $\widetilde{\Phi}_{1}\otimes\widetilde{\Phi}_{2}$ has
$\Phi$ as 1-restriction and that each of $\widetilde{\Phi}_{1}$ and
$\widetilde{\Phi}_{2}$ is trace preserving and respects
1,0-states. Hence, $\Phi$ is an LSP channel.
\end{proof}
\section{Subspace local channels}
\label{seSL}
In this section we turn to the SL channels in general and prove
explicit expressions to generate all SL channels.  It is shown that
the set of SL channels decomposes into a union of four disjoint
classes of channels, of which the LSP channels forms one such class.
\begin{Lemma}
\label{lenn}
Let $\phi$ be a CPM such that
$\Tr(P_{t1}\phi(|\psi_{s1}\rangle\langle\psi_{s1}|)) = 0$,
$\forall|\psi_{s1}\rangle\in\mathcal{H}_{s1}$ and
$\Tr(P_{t1}\phi(|\psi_{s2}\rangle\langle\psi_{s2}|)) = 0$,
$\forall|\psi_{s2}\rangle\in\mathcal{H}_{s2}$. Then
$P_{t2}\phi(|\psi\rangle\langle\psi|)P_{t2} =
\phi(|\psi\rangle\langle\psi|)$,
$\forall|\psi\rangle\in\mathcal{H}_{S}$.
\end{Lemma}
\begin{proof}
Let $\{V_{k}\}_{k}$ be an arbitrary Kraus representation of $\phi$.
According to lemma 1 in \cite{ref1}, the condition
$\Tr(P_{t1}\phi(|\psi_{s1}\rangle\langle\psi_{s1}|)) = 0$,
$\forall|\psi_{s1}\rangle\in\mathcal{H}_{s1}$ implies
$P_{t1}V_{k}P_{s1}=0$. The second condition similarly leads to
$P_{t1}V_{k}P_{s2}=0$. By combining these one obtains $P_{t1}V_{k}=0$,
which implies $P_{t2}V_{k} = V_{k}$, from which the lemma follows.
\end{proof}
\begin{Lemma}
\label{leca}
Let $\widetilde{\Phi}_{1}$ be a trace preserving CPM with source
$F_{01}(\mathcal{H}_{s1})$ and target $F_{01}(\mathcal{H}_{t1})$ and
let $\widetilde{\Phi}_{2}$ be a trace preserving CPM with source
$F_{01}(\mathcal{H}_{s2})$ and target $F_{01}(\mathcal{H}_{t2})$. If
$\widetilde{\Phi}_{1}\otimes\widetilde{\Phi}_{2}$ respects 1-states,
then exactly one of the following four cases is true
\begin{itemize}
\item 
$\widetilde{\Phi}_{1}$ is SP from
$(F_{0}(\mathcal{H}_{s1}),F_{1}(\mathcal{H}_{s1}))$ to
$(F_{0}(\mathcal{H}_{t1}),F_{1}(\mathcal{H}_{t1}))$ and \newline
\noindent $\widetilde{\Phi}_{2}$ is SP from 
$(F_{0}(\mathcal{H}_{s2}),F_{1}(\mathcal{H}_{s2}))$ to 
$(F_{0}(\mathcal{H}_{t2}),F_{1}(\mathcal{H}_{t2}))$.
\item 
$\widetilde{\Phi}_{1}$ is SP from
$(F_{0}(\mathcal{H}_{s1}),F_{1}(\mathcal{H}_{s1}))$ to
$(F_{1}(\mathcal{H}_{t1}),F_{0}(\mathcal{H}_{t1}))$ and \newline
\noindent $\widetilde{\Phi}_{2}$ is SP from 
$(F_{0}(\mathcal{H}_{s2}),F_{1}(\mathcal{H}_{s2}))$ to 
$(F_{1}(\mathcal{H}_{t2}),F_{0}(\mathcal{H}_{t2}))$.
\item $\widetilde{\Phi}_{1}(\widetilde{Q}_{1}) = |\widetilde{0}_{t1}\rangle
\langle \widetilde{0}_{t1}|\Tr(\widetilde{Q}_{1}),\quad 
P_{\widetilde{t2}:1}\widetilde{\Phi}_{2}(\widetilde{Q}_{2}) 
P_{\widetilde{t2}:1} = \widetilde{\Phi}_{2}(\widetilde{Q}_{2})$,
\item $P_{\widetilde{t1}:1}\widetilde{\Phi}_{1}(\widetilde{Q}_{1})
P_{\widetilde{t1}:1} = \widetilde{\Phi}_{1}(\widetilde{Q}_{1}),\quad  
\widetilde{\Phi}_{2}(\widetilde{Q}_{2}) = |\widetilde{0}_{t2}\rangle
\langle \widetilde{0}_{t2}|\Tr(\widetilde{Q}_{2})$,
\end{itemize}
for all $\widetilde{Q}_{1}\in\mathcal{L}(F_{01}(\mathcal{H}_{s1}))$
and for all
$\widetilde{Q}_{2}\in\mathcal{L}(F_{01}(\mathcal{H}_{s2}))$.
\end{Lemma}
\begin{proof}
Since $\widetilde{\Phi}_{1}\otimes\widetilde{\Phi}_{2}$ respects
1-states, it holds, by definition
\begin{eqnarray}
\label{eqcomp}
\Tr\left((P_{\widetilde{t1}:0}\otimes P_{\widetilde{t2}:0}+
P_{\widetilde{t1}:1}\otimes P_{\widetilde{t2}:1})
[ \widetilde{\Phi}_{1}\otimes\widetilde{\Phi}_{2}]
(|\widetilde{\psi}\rangle\langle\widetilde{\psi}|) \right)= 0,\nonumber \\ 
\forall |\widetilde{\psi}\rangle\in 
F_{0}(\mathcal{H}_{s1})\otimes 
F_{1}(\mathcal{H}_{s2})\oplus F_{1}(\mathcal{H}_{s1})
\otimes F_{0}(\mathcal{H}_{s2}).
\end{eqnarray} 
Equation (\ref{eqcomp}) implies the following conditions:
\begin{eqnarray}
\label{c1}
\Tr(P_{\widetilde{t1}:0} 
\widetilde{\Phi}_{1}(|\widetilde{\psi}_{0}^{s1}\rangle
\langle\widetilde{\psi}_{0}^{s1} |))
\Tr( P_{\widetilde{t2}:0}\widetilde{\Phi}_{2}
(|\widetilde{\psi}_{1}^{s2}\rangle\langle\widetilde{\psi}_{1}^{s2} |)) = 0,
 \nonumber\\ \quad \forall |\widetilde{\psi}_{0}^{s1}\rangle\in 
F_{0}(\mathcal{H}_{s1}),\quad \forall |\widetilde{\psi}_{1}^{s2}\rangle \in 
F_{1}(\mathcal{H}_{s2}), 
\end{eqnarray}
\begin{eqnarray}
\label{c2}
\Tr(P_{\widetilde{t1}:1}\widetilde{\Phi}_{1}(|\widetilde{\psi}_{0}^{s1}\rangle
\langle\widetilde{\psi}_{0}^{s1} |))\Tr(P_{\widetilde{t2}:1} 
\widetilde{\Phi}_{2}(|\widetilde{\psi}_{1}^{s2}\rangle
\langle\widetilde{\psi}_{1}^{s2} |)) = 0,\nonumber \\
 \quad \forall |\widetilde{\psi}_{0}^{s1}\rangle\in F_{0}(\mathcal{H}_{s1}),
\quad \forall |\widetilde{\psi}_{1}^{s2}\rangle \in F_{1}(\mathcal{H}_{s2}), 
\end{eqnarray}
\begin{eqnarray}
\label{c3}
\Tr( P_{\widetilde{t1}:0}\widetilde{\Phi}_{1}
(|\widetilde{\psi}_{1}^{s1}\rangle\langle\widetilde{\psi}_{1}^{s1} |))
\Tr(P_{\widetilde{t2}:0} 
\widetilde{\Phi}_{2}(|\widetilde{\psi}_{0}^{s2}\rangle
\langle\widetilde{\psi}_{0}^{s2} |)) = 0, \nonumber\\ 
\quad \forall |\widetilde{\psi}_{1}^{s1}\rangle\in 
F_{1}(\mathcal{H}_{s1}),\quad \forall |\widetilde{\psi}_{0}^{s2}\rangle \in 
F_{0}(\mathcal{H}_{s2}), 
\end{eqnarray}
\begin{eqnarray}
\label{c4}
\Tr(P_{\widetilde{t1}:1} \widetilde{\Phi}_{1}
(|\widetilde{\psi}_{1}^{s1}\rangle\langle\widetilde{\psi}_{1}^{s1} |))
\Tr(P_{\widetilde{t2}:1}
 \widetilde{\Phi}_{2}(|\widetilde{\psi}_{0}^{s2}\rangle
\langle\widetilde{\psi}_{0}^{s2} |)) = 0,\nonumber \\ 
\quad \forall |\widetilde{\psi}_{1}^{s1}\rangle\in 
F_{1}(\mathcal{H}_{s1}),\quad \forall |\widetilde{\psi}_{0}^{s2}\rangle 
\in F_{0}(\mathcal{H}_{s2}). 
\end{eqnarray}
If condition (\ref{c1}) is to be fulfilled then either
\begin{equation}
\label{eq1}
\Tr( P_{\widetilde{t1}:0} \widetilde{\Phi}_{1}
(|\widetilde{\psi}_{0}^{s1}\rangle\langle\widetilde{\psi}_{0}^{s1} |)) = 0,
\quad \forall |\widetilde{\psi}_{0}^{s1}\rangle\in F_{0}(\mathcal{H}_{s1}),
\end{equation}
or
\begin{equation}
\label{eq2}
\Tr(P_{\widetilde{t2}:0} \widetilde{\Phi}_{2}
(|\widetilde{\psi}_{1}^{s2}\rangle\langle\widetilde{\psi}_{1}^{s2} |)) = 0,
\quad \forall |\widetilde{\psi}_{1}^{s2}\rangle \in F_{1}(\mathcal{H}_{s2}). 
\end{equation}
First suppose (\ref{eq1}) is true. By the assumption that
$\widetilde{\Phi}_{1}$ is trace preserving follows
\begin{equation}
\label{eq3}
\Tr(P_{\widetilde{t1}:1} \widetilde{\Phi}_{1}
(|\widetilde{\psi}_{0}^{s1}\rangle\langle\widetilde{\psi}_{0}^{s1} |)) = 1,
\quad \forall\,\,\textrm{normalized}\,\,
 |\widetilde{\psi}_{0}^{s1}\rangle\in F_{0}(\mathcal{H}_{s1}).
\end{equation}
Equation (\ref{eq3}) together with condition (\ref{c2}) imply
\begin{equation}
\label{eq4}
\Tr( P_{\widetilde{t2}:1}\widetilde{\Phi}_{2}
(|\widetilde{\psi}_{1}^{s2}\rangle\langle\widetilde{\psi}_{1}^{s2} |)) = 0,
\quad \forall |\widetilde{\psi}_{1}^{s2}\rangle \in F_{1}(\mathcal{H}_{s2}). 
\end{equation}
Since $\widetilde{\Phi}_{2}$ is trace preserving it follows from
(\ref{eq4}) that
\begin{equation}
\label{eq4.5}
\Tr(P_{\widetilde{t2}:0} \widetilde{\Phi}_{2}
(|\widetilde{\psi}_{1}^{s2}\rangle\langle\widetilde{\psi}_{1}^{s2} |)) = 1,
\quad \forall\,\,\textrm{normalized}\,\,
 |\widetilde{\psi}_{1}^{s2}\rangle \in F_{1}(\mathcal{H}_{s2}), 
\end{equation}
which clearly contradicts (\ref{eq2}). Hence, if (\ref{eq1}) is true
then (\ref{eq2}) cannot be true.  If we on the other hand assume
(\ref{eq2}) to be true then, by a similar derivation as above, using
that $\widetilde{\Phi}_{2}$ is trace preserving and using (\ref{c2}),
we find
\begin{equation}
\label{eq6}
\Tr(P_{\widetilde{t1}:1} \widetilde{\Phi}_{1}
(|\widetilde{\psi}_{0}^{s1}\rangle\langle\widetilde{\psi}_{0}^{s1} |)) = 0,
\quad \forall |\widetilde{\psi}_{0}^{s1}\rangle\in F_{0}(\mathcal{H}_{s1}).
\end{equation}
Since $\widetilde{\Phi}_{1}$ is trace preserving, it is possible to
deduce a contradiction to (\ref{eq1}) from (\ref{eq6}). (Similarly as
(\ref{eq4.5}) is in contradiction with (\ref{eq2}).) To summarize:
either (\ref{eq1}) and (\ref{eq4}) are true (which we call case a.1),
or (\ref{eq2}) and (\ref{eq6}) are true (to be called case a.2), but
not both. By analogous reasoning it is possible to show, using
conditions (\ref{c3}) and (\ref{c4}), that either
\begin{eqnarray}
\label{cab1}
\Tr(P_{\widetilde{t1}:0} \widetilde{\Phi}_{1}
(|\widetilde{\psi}_{1}^{s1}\rangle\langle\widetilde{\psi}_{1}^{s1} |)) & = 0,
\quad \forall |\widetilde{\psi}_{1}^{s1}\rangle\in 
F_{1}(\mathcal{H}_{s1}),\nonumber\\
\Tr(P_{\widetilde{t2}:1} \widetilde{\Phi}_{2}
(|\widetilde{\psi}_{0}^{s2}\rangle\langle\widetilde{\psi}_{0}^{s2} |)) & = 0,
\quad \forall |\widetilde{\psi}_{0}^{s2}\rangle \in F_{0}(\mathcal{H}_{s2}),
\end{eqnarray}
(to be called case b.1) or
\begin{eqnarray}
\label{cab2}
\Tr(P_{\widetilde{t2}:0} \widetilde{\Phi}_{2}
(|\widetilde{\psi}_{0}^{s2}\rangle\langle\widetilde{\psi}_{0}^{s2} |)) & = 0,
\quad \forall |\widetilde{\psi}_{0}^{s2}\rangle \in
 F_{0}(\mathcal{H}_{s2}),\nonumber\\ 
\Tr(P_{\widetilde{t1}:1}\widetilde{\Phi}_{1}
(|\widetilde{\psi}_{1}^{s1}\rangle\langle\widetilde{\psi}_{1}^{s1} |)) & = 0,
\quad \forall |\widetilde{\psi}_{1}^{s1}\rangle\in F_{1}(\mathcal{H}_{s1})
\end{eqnarray}
(to be called called case b.2) is true, but not both.  The a-cases are
independent of the b-cases. Hence, there is in total four mutually
exclusive cases. These four cases are treated separately.

Case (a.1, b.1): According to lemma \ref{lenn}, equation (\ref{eq1})
together with the first equation in (\ref{cab1}) imply
$P_{\widetilde{t1}:1}\widetilde{\Phi}_{1}
(\widetilde{Q}_{1})P_{\widetilde{t1}:1} =
\widetilde{\Phi}_{1}(\widetilde{Q}_{1})$ for all
$\widetilde{Q}_{1}\in\mathcal{L}(F_{01}(\mathcal{H}_{s1}))$, which is
the first condition in the fourth case of the lemma.  Similarly
(\ref{eq4}) and the second equation in (\ref{cab1}) together with
lemma \ref{lenn} imply
$P_{\widetilde{t2}:0}\widetilde{\Phi}_{2}(\widetilde{Q}_{2})
P_{\widetilde{t2}:0}
= \widetilde{\Phi}_{2}(\widetilde{Q}_{2})$, for all
$\widetilde{Q}_{2}\in \mathcal{L}(F_{01}(H_{s2}))$. Since
$F_{0}(\mathcal{H}_{t2})$ is a one-dimensional space and since
$\widetilde{\Phi}_{2}$ is trace preserving, it follows that
$\widetilde{\Phi}_{2}$ can be written as
$\widetilde{\Phi}_{2}(\widetilde{Q}_{2}) =
|\widetilde{0}_{t2}\rangle\langle
\widetilde{0}_{t2}|\Tr(\widetilde{Q}_{2})$, which is the second
condition in the fourth case of the lemma.

Case (a.2, b.2): By very similar calculations as in case (a.1, b.1),
it follows that case (a.2, b.2) leads to the third case in the lemma.

Case (a.1, b.2): Equation (\ref{eq1}) together with the second
equation in (\ref{cab2}) imply that $\widetilde{\Phi}_{1}$ is SP from
$(F_{0}(\mathcal{H}_{s1}),F_{1}(\mathcal{H}_{s1}))$ to
$(F_{1}(\mathcal{H}_{t1}),F_{0}(\mathcal{H}_{t1}))$. Likewise
(\ref{eq4}) together with the first equation in (\ref{cab2}) imply
that $\widetilde{\Phi}_{2}$ is SP from
$(F_{0}(\mathcal{H}_{s2}),F_{1}(\mathcal{H}_{s2}))$ to
$(F_{1}(\mathcal{H}_{t2}),F_{0}(\mathcal{H}_{t2}))$. Hence, case (a.1,
b.2) implies the second case in the lemma.

Case (a.2, b.1): A very similar derivation as in case (a.1, b.2) gives
that case (a.2, b.1) implies the first case in the lemma.
\end{proof}
\begin{Proposition}
\label{classSL}
If $\Phi$ is subspace local from $(\mathcal{H}_{s1},\mathcal{H}_{s2})$
to $(\mathcal{H}_{t1},\mathcal{H}_{t2})$, then $\Phi$ belongs to
exactly one of the following classes
\begin{itemize}
\item[$C_{1}${\rm :}] $\Phi$ is LSP from 
$(\mathcal{H}_{s1},\mathcal{H}_{s2})$ to  
$(\mathcal{H}_{t1},\mathcal{H}_{t2})$.
\item[$C_{2}${\rm :}] There exists a density operator $\rho_{1}$ on 
$\mathcal{H}_{T}$ such that $P_{t1}\rho_{1}P_{t1} = \rho_{1}$, with
non-zero eigenvalues $\{\lambda^{1}_{n}\}_{n=1}^{N}$ and some
corresponding orthonormal set of eigenvectors
$\{|\rho_{n}^{1}\rangle\}_{n=1}^{N}$
$(N\leq\dim(\mathcal{H}_{t1}))$. There exists a density operator
$\rho_{2}$ on $\mathcal{H}_{T}$ such that $P_{t2}\rho_{2}P_{t2} =
\rho_{2}$, with non-zero eigenvalues $\{\lambda^{2}_{m}\}_{m=1}^{M}$
and some corresponding orthonormal set of eigenvectors
$\{|\rho_{m}^{2}\rangle\}_{m=1}^{M}$
$(M\leq\dim(\mathcal{H}_{t2}))$. There exists some matrices $C =
[C_{n,k}]_{n,k=1}^{N,K}$ and $D = [D_{m,l}]_{m,l=1}^{M,L}$, where $K =
\dim(\mathcal{H}_{s1})$, $L = \dim(\mathcal{H}_{s2})$, such that
$CC^{\dagger}\leq I_{N}$ and $DD^{\dagger}\leq I_{M}$, and
\begin{eqnarray}
\label{rt}
\fl \Phi(Q) = \rho_{1}\Tr(P_{s2}Q) + \rho_{2}\Tr(P_{s1}Q)\nonumber \\
+\sum_{nm}\sum_{lk}\langle s2,l|Q|s1,k\rangle
C_{n,k}D^{*}_{m,l}\sqrt{\lambda_{n}^{1}\lambda_{m}^{2}}|
\rho^{1}_{n}\rangle\langle\rho^{2}_{m}|
\\ +\sum_{nm}\sum_{lk}\langle s1,k|Q|s2,l\rangle
C^{*}_{n,k}D_{m,l}\sqrt{\lambda_{n}^{1}\lambda_{m}^{2}}|
\rho^{2}_{m}\rangle\langle\rho^{1}_{n}|,\quad\forall
Q\in\mathcal{L}(\mathcal{H}_{S}).\nonumber
\end{eqnarray}
\item[$C_{3}${\rm :}] There exists a density operator $\rho_{2}$ on 
$\mathcal{H}_{T}$ such that $P_{t2}\rho_{2}P_{t2} = \rho_{2}$, and a 
trace preserving CPM $\Phi_{2}$ with source $\mathcal{H}_{s2}$ and target 
$\mathcal{H}_{t2}$ such that 
\begin{equation}
\label{rt3}
\Phi(Q) = \rho_{2}\Tr(P_{s1}Q) + \Phi_{2}(Q),\quad\forall 
Q\in\mathcal{L}(\mathcal{H}_{S}).
\end{equation}
\item[$C_{4}${\rm :}] There exists a density operator $\rho_{1}$ on 
$\mathcal{H}_{T}$ such that $P_{t1}\rho_{1}P_{t1} = \rho_{1}$, and a
trace preserving CPM $\Phi_{1}$ with source $\mathcal{H}_{s1}$ and
target $\mathcal{H}_{t1}$ such that
\begin{equation}
\label{rt4}
\Phi(Q) = \rho_{1}\Tr(P_{s2}Q) + \Phi_{1}(Q),\quad\forall 
Q\in\mathcal{L}(\mathcal{H}_{S}).
\end{equation}
\end{itemize}
If a trace preserving CPM $\Phi$ fulfills the conditions of one of
these cases, then $\Phi$ is subspace local from
$(\mathcal{H}_{s1},\mathcal{H}_{s2})$ to
$(\mathcal{H}_{t1},\mathcal{H}_{t2})$.
\end{Proposition}
\begin{proof}
Since $\Phi$ is subspace local there exists, according to proposition
\ref{reduprod}, a product channel
$\widetilde{\Phi}_{1}\otimes\widetilde{\Phi}_{2}$ of trace preserving
CPMs, which has $\Phi$ as its 1-restriction, and is such that lemma
\ref{leca} is applicable. The strategy of this proof is the
following. For each of the four cases in lemma \ref{leca} it is shown
that one of the cases $C_{1}$, $C_{2}$, $C_{3}$, or $C_{4}$ is
implied. This is done by taking the 1-restriction of
$\widetilde{\Phi}_{1}\otimes\widetilde{\Phi}_{2}$ for each case in
lemma \ref{leca}. Moreover, for each of the cases $C_{j}$ the opposite
implication, stated in the end of the proposition, is proved.

We begin with the first case of lemma \ref{leca}.  It is
straightforward to convince oneself that the first case in lemma
\ref{leca} corresponds to LSP channels. LSP channels are SL channels,
hence the opposite implication stated in the proposition is proved.

Consider the second case in lemma \ref{leca}. Since
$\widetilde{\Phi}_{1}$ is SP from
$(F_{0}(\mathcal{H}_{s1}),F_{1}(\mathcal{H}_{s1}))$ to
$(F_{1}(\mathcal{H}_{t1}),F_{0}(\mathcal{H}_{t1}))$, it is an SP
gluing of two trace preserving CPMs $\widetilde{\Delta}_{a}$,
$\widetilde{\Delta}_{b}$ (see \cite{ref3} proposition
1). $\widetilde{\Delta}_{a}$ with source $F_{0}(\mathcal{H}_{s1})$ and
target $F_{1}(\mathcal{H}_{t1})$, $\widetilde{\Delta}_{b}$ with source
$F_{1}(\mathcal{H}_{s1})$ and target $F_{0}(\mathcal{H}_{t1})$. The
source space $F_{0}(\mathcal{H}_{s1})$ of $\widetilde{\Delta}_{a}$ is
one-dimensional. One can realize that if such a map is to be a trace
preserving CPM then it can be written
$\widetilde{\Delta}_{a}(\widetilde{Q}_{1})=
\widetilde{\rho}_{1}\langle
\widetilde{0}_{s1}|\widetilde{Q}_{1}|\widetilde{0}_{s1}\rangle$ for
all $\widetilde{Q}_{1}\in\mathcal{L}(F_{0}(\mathcal{H}_{s1}))$, where
$\widetilde{\rho}_{1}$ is a density operator on
$F_{1}(\mathcal{H}_{t1})$. $\widetilde{\Delta}_{b}$ on the other hand,
has the one-dimensional target space $F_{0}(\mathcal{H}_{t1})$. The
only possibility for such a map to be a trace preserving CPM is
$\widetilde{\Delta}_{b}(\widetilde{Q}_{1}) =
|\widetilde{0}_{t1}\rangle\langle
\widetilde{0}_{t1}|\Tr(\widetilde{Q}_{1})$ for all
$\widetilde{Q}_{1}\in \mathcal{L}(F_{1}(\mathcal{H}_{s1}))$. We have
that $K=\dim(F_{1}(\mathcal{H}_{s1}))=\dim(\mathcal{H}_{s1})$.  Let
$\{\lambda^{1}_{n}\}_{n=1}^{N}$ be the non-zero eigenvalues and
$\{|\widetilde{\rho}^{1}_{n}\rangle\}_{n=1}^{N}$ a corresponding set
of orthonormal eigenvectors of $\widetilde{\rho}_{1}$. Note that
$N\leq\dim(F_{1}(\mathcal{H}_{1}))=\dim(\mathcal{H}_{t1})$. The CPM
$\widetilde{\Delta}_{a}$ has a linearly independent Kraus
representation \cite{ref1} on the form
$\{\sqrt{\lambda^{1}_{n}}|\widetilde{\rho}^{1}_{n}\rangle\langle
\widetilde{0}_{s1}|\}_{n}$.  Likewise
$\{|\widetilde{0}_{t1}\rangle\langle \widetilde{s1},k|\}_{k=1}^{K}$ is
a linearly independent Kraus representation of
$\widetilde{\Delta}_{b}$. Since $\widetilde{\Phi}_{1}$ is an SP gluing
of $\widetilde{\Delta}_{a}$ and $\widetilde{\Delta}_{b}$ it follows,
by proposition 3 in \cite{ref3}, that $\widetilde{\Phi}_{1}$ can be
written as
\begin{eqnarray}
\label{la1}
\fl\widetilde{\Phi}_{1}(\widetilde{Q}_{1}) = 
\widetilde{\rho}_{1}
\langle\widetilde{0}_{s1}|\widetilde{Q}_{1}|\widetilde{0}_{s1}\rangle +
 |\widetilde{0}_{t1}\rangle\langle\widetilde{0}_{t1}|
\Tr(P_{\widetilde{s1}:1}\widetilde{Q}_{1}) \nonumber \\ 
 +\sum_{nk}
 C_{n,k}\sqrt{\lambda_{n}^{1}}|\widetilde{\rho}_{n}^{1}\rangle\langle
 \widetilde{0}_{s1}|\widetilde{Q}_{1}|\widetilde{s1},k\rangle
\langle\widetilde{0}_{t1}|\nonumber
 \\ +\sum_{nk}
 C_{n,k}^{*}|\widetilde{0}_{t1}\rangle
\langle\widetilde{s1},k|\widetilde{Q}_{1}|\widetilde{0}_{s1}\rangle
\langle\widetilde{\rho}_{n}^{1}
 |\sqrt{\lambda_{n}^{1}}
 ,\quad\forall\widetilde{Q}_{1}\in\mathcal{L}(F_{01}(\mathcal{H}_{s1})),
\end{eqnarray}
where the matrix $C = [C_{n,k}]_{n,k}$ fulfills the condition
$CC^{\dagger}\leq I_{N}$.  An analogous reasoning, but applied to
$\widetilde{\Phi}_{2}$, leads to
\begin{eqnarray}
\label{la2}
\fl \widetilde{\Phi}_{2}(\widetilde{Q}_{2}) = 
 \widetilde{\rho}_{2}\langle\widetilde{0}_{s2}|
\widetilde{Q}_{2}|\widetilde{0}_{s2}\rangle + 
|\widetilde{0}_{t2}\rangle
\langle\widetilde{0}_{t2}|\Tr(P_{\widetilde{s2}:1}\widetilde{Q}_{2})
\nonumber \\ 
 +\sum_{ml}
 D_{m,l}\sqrt{\lambda_{m}^{2}}|\widetilde{\rho}_{m}^{2}\rangle\langle
 \widetilde{0}_{s2}|\widetilde{Q}_{2}|\widetilde{s2},l\rangle
\langle\widetilde{0}_{t2}|\nonumber\\
 +\sum_{ml}
 D_{m,l}^{*}|\widetilde{0}_{t2}\rangle
\langle\widetilde{s2},l|\widetilde{Q}_{2}|\widetilde{0}_{s2}\rangle
\langle\widetilde{\rho}_{m}^{2}
 |\sqrt{\lambda_{m}^{2}},\quad
 \forall\widetilde{Q}_{2}\in\mathcal{L}(F_{01}(\mathcal{H}_{s2})),
\end{eqnarray}
where $\widetilde{\rho}_{2}$ is a density operator on
$F_{1}(\mathcal{H}_{t2})$, $\{\lambda_{m}^{2}\}_{m=1}^{M}$ the
non-zero eigenvalues and
$\{|\widetilde{\rho}_{m}^{2}\rangle\}_{m=1}^{M}$ a corresponding set
of orthonormal eigenvectors of $\widetilde\rho_{2}$. The matrix $D =
[D_{m,l}]_{m,l=1}^{M,L}$ fulfills the condition $DD^{\dagger}\leq
I_{M}$.  The 1-restriction of
$\widetilde{\Phi}_{1}\otimes\widetilde{\Phi}_{2}$ can be calculated
using (\ref{la1}), (\ref{la2}), and (\ref{reduc}). One may verify that
the result of the 1-restriction is case $C_{2}$, with
\begin{eqnarray}
\rho_{1} = M_{t1}^{\dagger}\widetilde{\rho}_{1}M_{t1},&
\quad \rho_{2} = M_{t2}^{\dagger}\widetilde{\rho}_{2}M_{t2},\nonumber\\
|\rho^{1}_{n}\rangle =
M_{t1}^{\dagger}|\widetilde{\rho}^{1}_{n}\rangle,&\quad
|\rho^{2}_{m}\rangle =
M_{t2}^{\dagger}|\widetilde{\rho}^{2}_{m}\rangle,\nonumber\\
|s1,k\rangle = M_{s1}^{\dagger}|\widetilde{s1},k\rangle,&\quad
|s2,l\rangle = M_{s2}^{\dagger}|\widetilde{s2},l\rangle.
\end{eqnarray}

To show that every channel of type $C_{2}$ is SL, it has to be shown
that every channel on the form (\ref{rt}) is an SL channel. Define
$\widetilde{\Phi}_{1}$ and $\widetilde{\Phi}_{2}$ by equations
(\ref{la1}) and (\ref{la2}), with
\begin{eqnarray}
\widetilde{\rho}_{1} = M_{t1}\rho_{1}M_{t1}^{\dagger},&
\quad\widetilde{\rho}_{2} = M_{t2}\rho_{2}M_{t2}^{\dagger},\nonumber\\
|\widetilde{\rho}^{1}_{n}\rangle =
M_{t1}|\rho^{1}_{n}\rangle,&\quad|\widetilde{\rho}^{2}_{m}\rangle =
M_{t2}|\rho^{2}_{m}\rangle,\nonumber \\ |\widetilde{s1},k\rangle =
M_{s1}|s1,k\rangle,&\quad|\widetilde{s2},l\rangle =
M_{s2}|s2,l\rangle.
\end{eqnarray}
By the assumed properties of $\rho_{1}$ and $\rho_{2}$ it follows that
$\widetilde{\Phi}_{1}\otimes\widetilde{\Phi}_{2}$ is trace preserving
and has $\Phi$ as its 1-restriction.

We next turn to the third case of lemma \ref{leca}. As a partial step
in the calculation of the 1-restriction one can use the following:
\begin{eqnarray*}
\fl \widetilde{\Phi}_{1}\otimes\widetilde{\Phi}_{2}
(|\widetilde{0}_{s1}\rangle\langle \widetilde{0}_{s1}|\otimes
|\widetilde{s2},l\rangle\langle \widetilde{s2},l'| ) &= 
|\widetilde{0}_{t1}\rangle\langle \widetilde{0}_{t1}|
\otimes\widetilde{\Phi}_{2}(|\widetilde{s2},l\rangle
\langle \widetilde{s2},l'|),\\
\fl \widetilde{\Phi}_{1}\otimes\widetilde{\Phi}_{2}
(|\widetilde{s1},k\rangle\langle \widetilde{s1},k'|\otimes
|\widetilde{0}_{s2}\rangle\langle \widetilde{0}_{s2}|)& = 
\delta_{kk'}|\widetilde{0}_{t1}\rangle\langle \widetilde{0}_{t1}|\otimes
\widetilde{\Phi}_{2}(|\widetilde{0}_{s2}\rangle\langle \widetilde{0}_{s2}|),
\end{eqnarray*}
\begin{equation*}
\fl\widetilde{\Phi}_{1}\otimes\widetilde{\Phi}_{2}
(|\widetilde{s1},k\rangle\langle \widetilde{0}_{s1} |
\otimes|\widetilde{0}_{s2}\rangle\langle \widetilde{s2},l|) = 0, \quad 
\widetilde{\Phi}_{1}\otimes\widetilde{\Phi}_{2}
(| \widetilde{0}_{s1}\rangle\langle\widetilde{s1},k' |
\otimes|\widetilde{s2},l\rangle\langle \widetilde{0}_{s2}|) = 0.
\end{equation*}
By taking the 1-restriction, case $C_{3}$ is found, with $\rho_{2} =
M_{T}^{\dagger}|\widetilde{0}_{t1}\rangle\langle
\widetilde{0}_{t1}|\otimes\widetilde{\Phi}_{2}
(|\widetilde{0}_{s2}\rangle\langle
\widetilde{0}_{s2}|)M_{T}$, and $\Phi_{2}$ defined as $\Phi_{2}(Q) =
M_{t2}^{\dagger}\widetilde{\Phi}_{2}(M_{s2}QM_{s2}^{\dagger})M_{t2}$. The
CPM $\Phi_{2}$, regarded as having source $\mathcal{H}_{s2}$ and
target $\mathcal{H}_{t2}$, is trace preserving.  The operator
$\rho_{2}$ is a density operator since $\widetilde{\Phi}_{2}$ is trace
preserving and
$P_{\widetilde{t2}:1}\widetilde{\Phi}_{2}(|\widetilde{0}_{s2}\rangle\langle
\widetilde{0}_{s2}|)P_{\widetilde{t2}:1} =
\widetilde{\Phi}_{2}(|\widetilde{0}_{s2}\rangle\langle
\widetilde{0}_{s2}|)$. From this also follows that
$P_{t2}\rho_{2}P_{t2} = \rho_{2}$.

To show that CPMs of the form (\ref{rt3}) are SL channels, define
$\widetilde{\Phi}_{1}$ and $\widetilde{\Phi}_{2}$ by
\begin{eqnarray}
\fl\widetilde{\Phi}_{1}(\widetilde{Q}_{1}) = 
|\widetilde{0}_{t1}\rangle\langle \widetilde{0}_{t1}|
\Tr(\widetilde{Q}_{1}),\quad\forall \widetilde{Q}_{1}\in 
\mathcal{L}(F_{01}(\mathcal{H}_{s1})),\nonumber\\
\fl \widetilde{\Phi}_{2}(\widetilde{Q}_{2})  =
 M_{t2}\Phi_{2}(M_{s2}^{\dagger}\widetilde{Q}_{2}M_{s2})M_{t2}^{\dagger} +
 \widetilde{\rho}_{2}\Tr(P_{\widetilde{s2}:0}\widetilde{Q}_{2}) ,\quad
\forall \widetilde{Q}_{2}\in \mathcal{L}(F_{01}(\mathcal{H}_{s2})), 
\end{eqnarray} 
where $\widetilde{\rho}_{2} = M_{t2}\rho_{2}M_{t2}^{\dagger}$. One can
show that $\widetilde{\Phi}_{1}\otimes\widetilde{\Phi}_{2}$ is trace
preserving and has $\Phi$ as its 1-restriction. (Hence, it respects
1-states. See lemma \ref{retr}). Hence, $\Phi$ is an SL channel.

By reasoning analogous to the third case, one finds that the fourth
case of lemma \ref{leca} implies $C_{4}$. Likewise, channels of the
form (\ref{rt4}) can be shown to be SL channels.
\end{proof}

As a quite direct consequence of proposition \ref{classSL} we have the
following corollary.
\begin{Corollary}
\label{sim}
Let $\Phi$ be an SL channel from $(\mathcal{H}_{s1},\mathcal{H}_{s2})$
to $(\mathcal{H}_{t1},\mathcal{H}_{t2})$. The following gives
necessary and sufficient conditions for $\Phi$ to belong to one of the
four classes of SL channels given in proposition (\ref{classSL}).
\begin{itemize}
\item[$C_{1}${\rm :}] $\Tr(P_{t1}\Phi(Q))= \Tr(P_{s1}Q),\quad
\forall Q\in\mathcal{L}(\mathcal{H}_{S})$.
\item[$C_{2}${\rm :}] $\Tr(P_{t1}\Phi(Q))= \Tr(P_{s2}Q),\quad
\forall Q\in\mathcal{L}(\mathcal{H}_{S})$.
\item[$C_{3}${\rm :}] $\Tr(P_{t2}\Phi(Q))= \Tr(Q),\quad
\forall Q\in\mathcal{L}(\mathcal{H}_{S})$.
\item[$C_{4}${\rm :}] $\Tr(P_{t1}\Phi(Q))= \Tr(Q),\quad
\forall Q\in\mathcal{L}(\mathcal{H}_{S})$.
\end{itemize}
\end{Corollary}
In words this corollary says that the four types of SL channels can be
classified according to how they handle the probability weights on the
two locations. The LSP channels preserve the probability weight on
each location. Class $C_{2}$ swaps the probability weights. Class
$C_{3}$ and $C_{4}$ concentrate the probability into one of the two
locations. Hence, the LSP channels are the only channels that do not
redistribute the probability weights between the two locations.

At first sight it may seem surprising that a channel, which we claim to
act locally, should have the power to redistribute the probability
weights between the two locations. However, seen from the point of view of
second quantization, this is not that surprising. We may take the
channels of type $3$ as an example. At location $1$ the channel
$\widetilde{\Phi}_{1}$ acts by returning the vacuum state
$|\widetilde{0}_{t1}\rangle$, no matter the input state. At the other
location, channel $\widetilde{\Phi}_{2}$ maps single-particle states to
single-particle states, and maps the vacuum state to a fixed
single-particle state. As seen, the local particle number is not
conserved, but the removal of the particle at location $1$ is
compensated for at location $2$, where the vacuum state is mapped to a
single-particle state. The two channels act independently of each
other, but together they act as if a particle was `transferred' from
location $1$ to location $2$.

The following proposition shows that the LSP channels can be
characterized as the subspace local SP channels.
\begin{Proposition}
\label{intersect}
The intersection of the set of SP channels from
$(\mathcal{H}_{s1},\mathcal{H}_{s2})$ to
$(\mathcal{H}_{t1},\mathcal{H}_{t2})$ and the set of SL channels from
$(\mathcal{H}_{s1},\mathcal{H}_{s2})$ to
$(\mathcal{H}_{t1},\mathcal{H}_{t2})$, is the set of LSP channels from
$(\mathcal{H}_{s1},\mathcal{H}_{s2})$ to
$(\mathcal{H}_{t1},\mathcal{H}_{t2})$.
\end{Proposition}
\begin{proof}
We first prove that $SP \supset LSP$. By corollary \ref{sim} every LSP
channel fulfills $\Tr(P_{t1}\Phi(Q))=\Tr(P_{s1}Q)$ for all
$Q\in\mathcal{L}(\mathcal{H}_{S})$. Hence, by proposition 4 in
\cite{ref1} it follows that $\Phi$ is SP. By construction $SL\supset
LSP$. By combining these inclusions it follows that $SL\cap SP \supset
LSP$. It remains to show the opposite inclusion. It is sufficient to
show that the last three families of SL channels, described in
proposition \ref{classSL}, cannot be SP channels. By combining
proposition 4 in \cite{ref1} and corollary \ref{sim} one finds that
none of the last three types of SL channels can be SP.
\end{proof}

The following proposition provides a `composition table' for SL
channels and shows that a composition of two SL channels is again an
SL channel, which fits with the intuitive notion of locally realizable
channels. Let $\mathcal{H}_{R}
=\mathcal{H}_{r1}\oplus\mathcal{H}_{r2}$ be finite-dimensional. Assume
that $\mathcal{H}_{r1}$ and $\mathcal{H}_{r2}$ are at least
one-dimensional.
\begin{Proposition}
\label{compoSLISL}
If a CPM $\Phi_{a}$ is SL from $(\mathcal{H}_{s1},\mathcal{H}_{s2})$
to $(\mathcal{H}_{t1},\mathcal{H}_{t2})$ and if a CPM $\Phi_{b}$ is SL
from $(\mathcal{H}_{t1},\mathcal{H}_{t2})$ to
$(\mathcal{H}_{r1},\mathcal{H}_{r2})$, then $\Phi_{b}\circ\Phi_{a}$ is
SL from $(\mathcal{H}_{s1},\mathcal{H}_{s2})$ to
$(\mathcal{H}_{r1},\mathcal{H}_{r2})$.

If moreover $\Phi_{a}$ belongs to class $C_{i}$ and $\Phi_{b}$ belongs
to class $C_{j}$, then $\Phi_{b}\circ\Phi_{a}$ belongs to class
$C_{k}$ according to the following rules:
\begin{equation}
\label{1j}
C_{1}\circ C_{j} \subset C_{j},\quad C_{j}\circ C_{1} \subset
C_{j}\quad j=1,\ldots,4
\end{equation}
\begin{equation}
\label{34j}
C_{i}\circ C_{j}\subset C_{i}\quad i =3,4\quad j =1,\ldots,4
\end{equation}
\begin{equation}
\label{22}
C_{2}\circ C_{2}\subset C_{1},\quad C_{2}\circ C_{3}\subset
C_{4},\quad C_{2}\circ C_{4}\subset C_{3}
\end{equation}
\end{Proposition}
\begin{proof}
Since $\Phi_{a}$ and $\Phi_{b}$ are SL there exists trace preserving
product channels $\widetilde{\Phi}_{a1}\otimes\widetilde{\Phi}_{a2}$
and $\widetilde{\Phi}_{b1}\otimes\widetilde{\Phi}_{b2}$, with
$\Phi_{a}$ respectively $\Phi_{b}$ as 1-restrictions. Hence
$\Phi_{a}(Q) =
M_{T}^{\dagger}[\widetilde{\Phi}_{a1}\otimes
\widetilde{\Phi}_{a2}](M_{S}QM_{S}^{\dagger})M_{T}$
and $\Phi_{b}(Q) =
M_{R}^{\dagger}[\widetilde{\Phi}_{b1}\otimes
\widetilde{\Phi}_{b2}](M_{T}QM_{T}^{\dagger})M_{R}$. Using
$M_{T}M_{T}^{\dagger}=P_{\widetilde{T}:1}$ it follows that
\begin{eqnarray}
\Phi_{b}\circ\Phi_{a}(Q) & = 
M_{R}^{\dagger}[\widetilde{\Phi}_{b1}\otimes\widetilde{\Phi}_{b2}]
(P_{\widetilde{T}:1} 
[\widetilde{\Phi}_{a1}\otimes\widetilde{\Phi}_{a2}](M_{S}QM_{S}^{\dagger}) 
P_{\widetilde{T}:1})M_{R}\nonumber\\ 
 & =
 M_{R}^{\dagger}[(\widetilde{\Phi}_{b1}\circ\widetilde{\Phi}_{a1})\otimes
(\widetilde{\Phi}_{b2}\circ\widetilde{\Phi}_{a1})](M_{S}QM_{S}^{\dagger})M_{R}.
\end{eqnarray}
The last equality holds according to lemma \ref{lere}, since
$\widetilde{\Phi}_{a1}\otimes\widetilde{\Phi}_{a2}$ respects
1-states. Hence,
$(\widetilde{\Phi}_{b1}\circ\widetilde{\Phi}_{a1})\otimes
(\widetilde{\Phi}_{b2}\circ\widetilde{\Phi}_{a1})$ has
$\Phi_{b}\circ\Phi_{a}$ as 1-restriction. Hence,
$\Phi_{b}\circ\Phi_{a}$ is an SL channel. The rest of the proposition
follows from corollary \ref{sim}.
\end{proof}
\section{Discussion}
\label{dis}
Here we discuss some conceptual aspects of the SL channels in general.
Imagine for a moment that we live in a universe where all particles
are distinguishable and that none of these can be annihilated and
recreated again. Suppose we have two boxes and one particle. These two
boxes are located far away from each other and the particle can be in
any state of superposition or mixture of being localized in the
boxes. Consider general operations on the state of this
particle. Hence, we are considering channels with identical source and
target space $\mathcal{H}_{s1}\oplus\mathcal{H}_{s2}$, where
$\mathcal{H}_{s1}$, $\mathcal{H}_{s2}$ represent the pure localized
states on the two boxes. What subspace local channels on
$(\mathcal{H}_{s1},\mathcal{H}_{s2})$, of this type, are possible to
perform subspace locally? With the restrictions assumed for this toy
universe and with the assumed setup, only the LSP channels are
possible to perform subspace locally; the reason being that the other
three classes of SL channels do not conserve the local particle
number. If there are identical particles, a particle can be removed
(or perhaps annihilated) from the first box, and an identical particle
can be inserted (or perhaps created) at the second box. In our toy
universe, however, there is no identical particle to be inserted or
created in the other box. Hence, the only possibility to realize such
a channel would be to physically transport the particle from one box
to the other, which cannot reasonably be called a `local'
operation. Far from being clearcut, this example suggests that there
is a connection between questions of locality and non-locality of
quantum channels and the existence of identical particles. At least
from the point of view presented here, a universe with identical
particles seems more `allowing' than a universe without.

From the discussion above one may be tempted to make the conclusion
that in a universe with only distinguishable particles, the only SL
channels that are possible to realize subspace locally, are the LSP
channels. This is however a too rapid conclusion, as the following
examples show. Consider again the two-box system described above, but
this time with two distinguishable particles instead of one. We call
them the S-particle and the T-particle.  The state space of the S
particle on the two boxes is
$\mathcal{H}_{S}=\mathcal{H}_{s1}\oplus\mathcal{H}_{s2}$, where
$\mathcal{H}_{s1}$ represents the pure states of particle S localized
in box $1$ and so on. Likewise $\mathcal{H}_{t1}$ and
$\mathcal{H}_{t2}$ represents the pure localized states of the
T-particle in the two boxes. We let the T-particle be in some fixed
initial state $\rho_{T}$ and ask how the state of the S-particle
affects the state of the T-particle, if the particles initially are in
a product state $\rho_{S}\otimes\rho_{T}$. To create examples of SL
channels of type $C_{4}$, we let the T-particle be localized in box
$1$ ($P_{t1}\rho_{T}P_{t1}= \rho_{T}$).  The most simple example of a
SL mapping of type $C_{4}$ is the extreme case of no interaction
between the two particles. In that case one obtains the map
$\Phi(\rho_{S}) = \Tr_{S}(\rho_{S}\otimes\rho_{T}) = \rho_{T}$. It is
possible, however, to create a bit less trivial examples. We assume
the S and T particle to interact only if they occupy the same box. It
seems reasonable to assume an Hamiltonian on the following form. For
the sake of simplicity we only consider interactions and have no
Hamiltonians for the particles themselves.
\begin{eqnarray}
\label{intr1}
H = & \sum_{k,k',n,n'}H_{k,n;k',n'}|s1,k\rangle\langle s1,k'|\otimes
|t1,n\rangle\langle t1,n'|\nonumber\\ & +
\sum_{l,l',m,m'}H_{l,m;l',m'}|s2,l\rangle\langle s2,l'|\otimes
|t2,m\rangle\langle t2,m'|.
\end{eqnarray}
Again we assume the initial state to be a product state and let the
T-particle be in a fixed localized state, and ask how the initial
S-state affects the T-state after some fixed (but arbitrary) elapse of
time $t$. The following channel answers this question: $\Phi(\rho_{S})
= \Tr_{S}(e^{-itH}\rho_{S}\otimes\rho_{T}e^{itH})$, for all density
operators $\rho_{S}$ on $\mathcal{H}_{S}$, where we let $\hbar=1$. It
is possible to deduce that $\Phi$ is an SL channel of type $C_{4}$, by
using $P_{t1}\rho_{T}P_{t1}= \rho_{T}$ and the structure of
\ref{intr1}. This is true irrespective of the choice of time $t$. Note
that the initial state $\rho_{T}$ is chosen to be local. Hence, in some
sense one does not expect it to be a `non-local resource'. As seen,
neither annihilation nor creation of particles has been assumed in
this construction.

We can conclude that if we consider state changes on one
and the same system, only the LSP channels are locally realizable in
this `gedanken universe', while if we consider mappings from one
system to another, also other SL channels are locally
realizable. Hence, in the latter case there is, in some sense, more
freedom. This suggest that the set of channels which are locally
realizable depends on the specific context.

Before ending this section we mention one further aspect on the
relation between the LSP channels and the rest of the SL channels. One
intuitively reasonable requirement for an operation to be local with
respect to two locations, is that it should be possible to compose it
out of two operations: one which operates on location 1, while doing
`nothing' on location 2, and the other operating on the second location
while doing nothing on the first. In case of locality with respect to
tensor product decomposition this is trivially satisfied, since
$\Phi_{1}\otimes\Phi_{2} = (I_{1}\otimes\Phi_{2})\circ(\Phi_{1}\otimes
I_{2})$. By the very construction of the here proposed definition of
subspace locality, all SL operations can be decomposed in this way, in
terms of operations on the second quantized spaces. This since any SL
channel corresponds to a product channel
$\widetilde{\Phi}_{1}\otimes\widetilde{\Phi}_{2}$ on the second
quantized spaces. However, the LSP channels do admit a much simpler
decomposition, directly in terms of the first quantized spaces. A
reasonable interpretation of ``operating on location 1 and doing
nothing on location 2'', is to have a channel which is a trace
preserving gluing \cite{ref3} of a channel on location 1 and an
identity CPM on location 2.
\begin{Proposition}
\label{LSPdeco}
A channel $\Phi$ is LSP from $(\mathcal{H}_{s1},\mathcal{H}_{s2})$ to
$(\mathcal{H}_{r1},\mathcal{H}_{r2})$, if and only if there exist
channels $\Phi_{a}$ and $\Phi_{b}$ such that $\Phi=
\Phi_{b}\circ\Phi_{a}$, where $\Phi_{a}$ is a trace preserving gluing
of a channel with source $\mathcal{H}_{s1}$ and target
$\mathcal{H}_{t1}$, and the identity CPM with source and target
$\mathcal{H}_{s2}$, and where $\Phi_{b}$ is a trace preserving gluing
of a channel with source $\mathcal{H}_{s2}$ and target
$\mathcal{H}_{t2}$, and the identity CPM with source and target
$\mathcal{H}_{t1}$
\end{Proposition}  

\begin{proof}
We begin with the ``if'' part.  $\Phi_{a}$ and $\Phi_{b}$ are both LSP
channels since a trace preserving gluing of a trace preserving CPM and
an identity CPM necessarily is an LSP channel (see proposition 8 in
\cite{ref3}). By proposition \ref{compoSLISL} it follows that
$\Phi=\Phi_{b}\circ\Phi_{a}$ is LSP.

For the ``only if'' part, let $\{V_{n}\}_{n}$, V, $\{W_{m}\}_{m}$, and
$W$, be the operators in proposition \ref{islekv}, with respect to the
LSP channel $\Phi$. Let $\Phi_{a}(Q)= \sum_{n}V_{n}QV_{n}^{\dagger} +
P_{s2}QP_{s2} + VQP_{s2} + P_{s2}QV^{\dagger}$, for all
$Q\in\mathcal{L}(\mathcal{H}_{S})$. Let $\Phi_{b}(Q)= P_{t1}QP_{t1}
+\sum_{m}W_{m}QW_{m}^{\dagger}+ P_{t1}QW^{\dagger} + WQP_{t1}$, for
all $Q\in
\mathcal{L}(\mathcal{H}_{t1}\oplus\mathcal{H}_{s2})$. Clearly
$\Phi=\Phi_{b}\circ\Phi_{a}$ and one can check that each of $\Phi_{a}$
and $\Phi_{b}$ are gluings of channels and identity CPMs, as stated in
the proposition.
\end{proof}

There are more questions that may be raised on the nature of the
concept of subspace locality and the here proposed way to formalize
it. More investigation is needed to settle which is the most
preferable definition of subspace locality in different
contexts. Further aspects of subspace locality are discussed in
\cite{ref3}.
\section{Summary}
\label{sum}
A definition of \emph{subspace locality} (SL) of quantum channels
(trace preserving completely positive maps) is proposed. The purpose
of this definition is to formulate conditions that channels have to
fulfill, if they are to act `locally', when the division in locations
naturally corresponds to an orthogonal decomposition of the Hilbert
space, rather than a tensor product decomposition. One example of such
a system is a particle in a two-path interferometer, where the total
Hilbert space of the particle can be decomposed into an orthogonal sum
of two Hilbert spaces, each representing pure localized states in one
of the paths. Given a quantum channel acting on the state of a
particle in the two paths, we wish to find some condition that the
channel has to fulfill, if it is to act `locally' on each path.
  
The here proposed definition of subspace locality is stated in terms
of occupation number representations of second quantizations of the
involved Hilbert spaces. It is used that in the occupation number
representation, a second quantization of an orthogonal sum of two
subspaces, is equivalent to a tensor product of the second
quantizations of each of the two subspaces. With respect to this
tensor product decomposition, the `usual' definition of a locally
acting channel as a product channel, is used. The consequences of this
choice of definition is investigated.

Under the restricting assumption that the first quantized state spaces
are all finite-dimensional, the here proposed definition of subspace
locality of quantum channels is reformulated in the original first
quantized state spaces. This gives expressions which make it possible
to explicitly generate all subspace local channels. Moreover, it is
shown that the set of all SL channels decomposes into four disjoint
families. One of these families, called \emph{local subspace
preserving} (LSP), is shown to be the intersection between the set of
SL channels and the set of subspace preserving channels
\cite{ref1}. Proposition \ref{islekv} gives an explicit construction
of all the LSP channels. Proposition \ref{classSL} provides explicit
expressions for all the four families of SL channels.  For a subclass
of the LSP channels a special form of construction in terms of a joint
unitary evolution with an ancilla system is proved.  

\ack I thank Erik Sj\"oqvist for many valuable comments and discussions on the
manuscript. I also thank Marie Ericsson for discussions which started
the train of thoughts leading to this investigation. Finally I thank
Osvaldo Goscinski for reading and commenting the text.

\end{document}